\documentclass[aps,prx,reprint,groupedaddress,longbibliography,footinbib,floatfix]{revtex4-1}
\usepackage{graphicx}
\usepackage{placeins}
\usepackage{color}
\usepackage{amsmath}
\usepackage{amsfonts}
\usepackage{bm}
\usepackage{amssymb}
\usepackage[T1]{fontenc}
\usepackage[english]{babel}
\usepackage{hyperref}
\usepackage{array}
\usepackage{braket}
\usepackage{ragged2e}
\usepackage{lipsum} 
\usepackage{float}
\usepackage{adjustbox}
\usepackage[bottom]{footmisc}
\usepackage{booktabs}
\usepackage{comment}

\addto\captionsenglish{}
\addto\captionsenglish{}

\begin{document}


\title{Topological transitions in fluid lipid vesicles: activation energy and force fields}

\author{Matteo Bottacchiari, Mirko Gallo, Marco Bussoletti}%
\author{Carlo Massimo Casciola}
\email{carlomassimo.casciola@uniroma1.it}
\affiliation{Department of Mechanical and Aerospace Engineering, \\ Sapienza Universit\`a di Roma, Rome, Italy}%

\date{\today}

\begin{abstract}
Lipid bilayers possess two extraordinary and apparently conflicting properties that make them essential  
barriers at the cellular and sub-cellular level: they are at the same time very stable and extremely plastic. Due to their plasticity, they can easily change both their shape and their topology. For example, erythrocytes are reshaped to flow into narrow capillaries, while the topological changes are required for cellular processes such as endo- and exocytosis or to enable neurotransmitters to cross the neural synapses. On the other hand, stability is a  necessary prerequisite for their barrier function and for the complex cellular clockworks. It is a fact that there is currently no theoretical model able to encompass the full-scale morphological modifications of lipid bilayer vesicles. The classical description is the celebrated Canham-Helfrich model which possesses undeniable merits except the ability to deal with topological transitions. Muscular approaches such as molecular dynamics, even in its coarse-grained versions, provide invaluable microscopic insight but, for the forthcoming years, will still not be able to afford the full-scale simulation of the morphological changes in large and giant unilamellar vesicles.  In this scenario, the purpose of the present paper is to produce and demonstrate an approach belonging to the class of phase-field models, which offers the potential to capture the dynamics of these full-scale lipid vesicles, and the subtle effects of global constraints induced by the Gauss-Bonnet theorem of differential geometry, and by area and volume conservation. From the technical point of view, the breakthrough consists of a novel formulation to include the contribution of the membrane Gaussian curvature in the phase-field free energy functional. It will be shown that the energy barrier that stabilizes the vesicle topology is largely due to the contribution of the Gaussian curvature, rather than to the membrane bending rigidity. The topological transition, occurring in the fusion/fission process of simple vesicles, is addressed with techniques borrowed from the statistical mechanics of rare events. The force fields capable of inducing the transition with minimal work expenditure are analyzed and found consistent with known mechanisms that operate at the biochemical level. This picture is consistent with the intuition that protein systems could have evolved in such a way as to minimize the work needed to induce the topological transition by following a minimal free energy path. The model correctly identifies the scales relevant to the two strongly asymmetric processes of vesicle fusion and fission, and provides predictions that are quantitatively consistent with experimental estimates on the energy barrier and on the strength of the forces exerted by the protein systems involved in topological transitions. Intriguingly, only quite general and macroscopic parameters are needed, namely bilayer thickness and its rigidity, suggesting that the topological change is a quite generic process.
\end{abstract}



\maketitle

\section{Introduction}\label{INTRO}

Biological membranes are formed by a fluid lipid bilayer which can be described not only from a biochemical point of view but also from a mechanical one. The most important example of fluid lipid membrane is the plasma membrane, which defines the boundary between the inside and outside of cells. Topological transitions of fluid lipid membranes are involved in most of the fundamental processes of cell life, like endocytosis and exocytosis. An example of topological transformation is the merging of two membranes. This is the case of vesicle-vesicle fusion or viral membrane fusion. Indeed, viruses enveloped by a lipid bilayer, such as HIV, Ebola virus, influenza, measles, rabies virus and SARS-CoV-2 can infect a target cell by fusion of their membrane with the cell plasma membrane. The aim of this fusion is to inject the viral genome into the cell \cite{Barrett2020ViralMF, Harrison2008ViralMF}. Viral infection of the cell can also occur via endocytosis, in which the plasma membrane undergoes fission to internalize the virus via an endosome. Therefore, another important topological change is membrane fission, which is also fundamental for cell division and therefore for life \cite{Carlton2020MembraneAO, Chernomordik2003ProteinlipidII}. Topological transitions of lipid membranes are of great interest not only in biology and biophysics but also in medicine and in the pharmaceutical industry. Indeed, lipid-based nanoparticles are used for drug delivery, offering many advantages including biocompatibility, bioavailability, self-assembly and payload flexibility. For example, liposomes, namely lipid bilayer vesicles with a diameter in the range of $25 \; \text{nm} \div 1 \; \mu \text{m}$, are able to carry both hydrophilic and hydrophobic drugs \cite{Tenchov2021LipidNL}. Micelles, closed lipid monolayers, are currently used in mRNA-vaccines against COVID-19 and many other lipid nanoparticle-mRNA applications are under clinical evaluation, e.g. for the treatment of cancer or genetic diseases \cite{Hou2021LipidNF}. Regardless of the specific application, all these nanoparticles are engineered to overcome many physiological barriers, some of which involve topological transitions of lipid membranes. For instance, a topological transformation is needed for the cellular uptake of drugs and others may be required in the cytosol, e.g., to enter the nucleus for DNA delivery or genome editing. Therefore, a better comprehension of the mechanisms underlying such barriers could be useful to enhance the delivery efficiency, especially moving toward precision medicine \cite{Mitchell2020EngineeringPN}.

Fluid lipid membranes can be mechanically described using a continuum approach initially introduced in \cite{Canham1970TheME, Helfrich1973ElasticPO}. Such a classical elastic perspective describes a membrane as a two-dimensional surface $\Gamma$ with an energy density depending on its principal curvatures. An expansion of this density up to the second order in curvatures leads to the Canham-Helfrich Hamiltonian:
\begin{equation}\label{Canham-Helfrich}
	E_{CH} = 2k \int_{\Gamma} (M-m)^2 \; dS + k_G \int_{\Gamma} G \; dS.
\end{equation}
Here, the first term on the right-hand side is the bending energy and the second is the Gaussian energy. $M$ is the mean curvature of the surface, $G$ its Gaussian curvature, $m$ a spontaneous mean curvature that the membrane tends to adopt in absence of external forces, and $k$ and $k_G$ are called bending rigidity and Gaussian curvature modulus, respectively. $k$ can be experimentally measured in different ways \cite{Dimova2014RecentDI}, whereas $k_G$ is more elusive due to the celebrated Gauss-Bonnet (GB) theorem, 
\begin{equation}\label{Gauss_Bonnet_theorem}
	\int_{\Gamma} G \; dS = 2 \pi \chi(\Gamma) - \int_{\partial \Gamma} k_g \; dl,
\end{equation}
where $\chi(\Gamma)$ is the Euler characteristic of $\Gamma$ and $k_g$ is the geodesic curvature of the surface boundary  $\partial \Gamma$. In the vesicles case, since they are compact surfaces without boundary, the line integral vanishes, and $\chi (\Gamma)$ becomes equal to $2 - 2g$, being $g$ the genus of the surface. Therefore, the Gaussian energy term remains constant as long as no topological transitions occur, leading to the aforementioned elusive behavior of $k_G$. A stability argument \cite{DESERNO_GEOMETRY} shows that $-2 < k_G/k < 0$ and in literature there is evidence \cite{KG1, KG2, KG3} that $k_G \approx -k$.  Hence, the Gaussian energy is expected to play a crucial role during topological transitions. 
Because of the scale invariance of the Canham-Helfrich free energy, for a given topology, vesicles shapes are dictated by their reduced volume $v = V/(\pi \, D_\text{ve}^3 / 6)$,  as well as by their reduced spontaneous curvature $m_{0} = m D_\text{ve}$, where $D_{ve} = \sqrt{A/\pi}$ is the characteristic lenght of the vesicle under consideration, having area $A$ and volume $V$. 

The Canham-Helfrich description is thought to hold for vesicles with a characteristic length $D_{ve} \ge 40 \; l_{me}$ \cite{Dimova2019TheGV}, being $l_{me}$ the lipid bilayer thickness, which is usually about $5 \; \text{nm}$; otherwise, higher-order terms in the energy density could make a significant contribution. Therefore, Eq.~(\ref{Canham-Helfrich}) describes vesicles larger than the ones simulated by means of coarse-grained molecular dynamics (MD) and dissipative particle dynamics (DPD), which have been the most widely used techniques for in silico studies of topological transitions to date. Indeed, these computer simulations, which take into account the molecular details of lipid bilayers, allow monitoring in time morphological changes of small liposomes with a size below $50 \; \text{nm}$ \cite{Smirnova2015FreeenergyCM}. For example, membrane adsorption of small solutes inducing budding and fission of nanoparticles has been simulated in \cite{Ghosh2021BuddingAF}, membrane fusion and drug delivery with carbon nanotube porins in \cite{Ho2021MembraneFA}, whereas the intermediate structures in which lipids are organized during fusion reactions have been investigated in a series of works including \cite{Marrink2003TheMO, Knecht2007MolecularDS, Shillcock2005TensioninducedFO, Grafmller2007PathwayOM, Smirnova2019ThermodynamicallyRP}. Theoretical description of these intermediates has been addressed, among others, in \cite{Kozlovsky2002LipidII, Efrat2007PointlikePA, Siegel2004TheGC, Siegel2008TheGC, Jackson2009MinimumMB} and the same has been done for membrane fission in \cite{Kozlovsky2003MembraneFM}. Membrane fusion and fission events have also been experimentally studied. For example, recently, controlled fission of cell-sized vesicles by low densities of membrane-bound proteins has been reported in \cite{Steinkhler2020ControlledDO}. Other examples of fission experiments can be found in \cite{Dbereiner1993BuddingAF, Avinoam2015EndocyticSM}, whereas, as regards fusion, merging of giant liposomes has been observed in \cite{Lei2003LipidBV, Fix2004ImagingSM, Haluska2006TimeSO}, the stalk intermediate in \cite{Aeffner2012EnergeticsOS}, and activation energies for small liposomes fusion events have been measured in \cite{Lee1998SecretoryAV, FranoisMartin2017LowEC} by means of kinetic analysis.

In this work, we numerically study fission and fusion events of Helfrich-sized vesicles. In order to achieve our purpose, we use the phase-field framework \cite{Lzaro2015PhasefieldTF, Jamet2008TowardAT}, which allows us not to track the interface and naturally handle topological transitions. A phase-field version of the bending energy term alone has been initially introduced in \cite{Du2004APF, Du2005ModelingTS, Du2006SimulatingTD}, leading to numerous applications like \cite{Campelo2006DynamicMA,Gu2014SimulatingVA, Gu2016ATP, Wang2008ModellingAS, Campelo2007ModelFC, Lzaro2014RheologyOR, Lzaro2019CollectiveBO, Barrio2020TheDO}. Furthermore, in \cite{Du2005RetrievingTI}, it has been pointed out that it is possible to retrieve topological information for phase-field models. Here, given its importance in topological transitions, we include the Gaussian energy term in the dynamics. This coupling has a rigorous mathematical derivation, which is reported in Section \ref{PFMD}. The new term numerically captures with high accuracy the energy jumps as expected by the GB theorem; see for comparison the recent paper \cite{Rueda2021gaussian}, where a different form of phase-field Gaussian energy was proposed to study pearling instabilities. Interestingly, Appendix~\ref{Gauss_Bonnet_appendix}, the GB theorem can be derived from our new phase-field energy term.

In this paper, by means of the string method \cite{Weinan2002StringMF}, a rare event technique, we compute a minimal energy pathway (MEP) between two spherical vesicles and a dumbbell-shaped one (see, e.g., \cite{ren2014wetting, lutsko2019crystals, gallo2020nucleation, magaletti2021water} for the string method applied to a phase-field approach). This configuration is of special interest for applications since most of the aforementioned experimental results refer to fusion/fission processes involving spherical vesicles. However, the approach we propose can be easily applied also to other vesicle configurations like, e.g., two spheres of different sizes, two non-spherical vesicles or to the limiting case of a sphere and a plane. The string method, coupled with the new phase-field model, provides the free energy barriers of the fusion and fission processes and the membranes configurations along the path. Moreover, since the phase-field extension of the Canham-Helfrich energy regularizes the behavior of the Gaussian term, we also compute the force fields needed to overcome these barriers. These forces are necessary to balance the elastic reaction arising from the bending and Gaussian energies, as well as from the membrane incompressibility. In the classical Canham-Helfrich approach, the computation of the bending forces would not have been so easy \cite{Guckenberger2017TheoryAA} and it would not have been clear how to include the Gaussian contribution during topological transitions. The computation of such forces will pave the way for exploring the effective mechanisms by which fusion and fission machinery work across the full scale of vesicles.

The paper is structured as follows: Section~\ref{PFMD} illustrates the main aspects of the underlying mathematics; this section is not essential for the discussion of the physical results. Section~\ref{Results} details the obtained results, which are discussed in Section~\ref{Discussion}; final remarks are reported in Section~\ref{Final Remarks}, whereas Section~\ref{Methods} treats the more technical aspects (numerical scheme, string method, and the computation of the force fields).

\section{Model derivation}\label{PFMD}

The classical Canham-Helfrich model relies on a sharp interface description of the membrane, which is treated as a (zero thickness) surface endowed with an energy density depending on the principal curvatures, Eq.~(\ref{Canham-Helfrich}). The model succeeds in describing many aspects of the vesicle dynamics but rules out the possibility of dealing with topological changes, unless unphysical surgical operations are conceived to cut and paste patches of the membrane.

A viable alternative to the sharp interface description is to employ a smooth function defined on a domain $\Omega$ -- the phase-field $\phi(\bm{x})$ -- that discriminates between the inner and the outer environment of the vesicle assuming the limiting values $\pm 1$ in the two regions. The $\phi(\bm{x}) = 0$ level set represents the membrane midsurface $\Gamma$. The transition between the two limiting values takes place in a narrow region whose width is controlled by a small parameter $\epsilon$ and can be associated with the thickness of the lipid bilayer. The main advantage of describing the membrane with such a field lies in the fact that it enables topological modifications of the membrane, allowing to address the problem of vesicle fusion and fission.

An energy $E[\phi]$ is associated with each field configuration and is such as to admit local minimizers of the form
\begin{equation}\label{ansatz}
\phi(\bm{x}) = f\bigg(\dfrac{d(\bm{x})}{\epsilon}\bigg),
\end{equation} 
where $d(\cdot)$ is the signed distance function from the membrane midsurface $\Gamma$.  We choose to define the signed distance such that $\bm{n}=\bm{\nabla} d$ computed on $\Gamma$ is equal to the inward-pointing unit normal to the vesicle. Setting $d^*(\bm{x}) = d(\bm{x}) / \epsilon$, we also require that $\lim_{d^* \to \pm \infty} \phi = \pm 1$ and $\phi = 0$ for $d=0$. Therefore, $\pm 1$ are the values for the stable phases of the inside and outside bulk and the level set $\phi = 0$ identifies the membrane midsurface. Physically, the energy functional should recover the Canham-Helfrich energy, (\ref{Canham-Helfrich}), in the limit of small width-to-vesicle-extension ratio.

The proposed expression for the phase-field free energy functional is
\begin{equation}\label{Energy}
E[\phi] = E_B[\phi] + E_G[\phi],
\end{equation}
with
\begin{equation}\label{Bending}
E_{B}[\phi] = k \, \dfrac{3}{4 \sqrt{2}} \, \epsilon \, \int_{\Omega} \Psi_{B}^2 \; dV,
\end{equation}
\begin{equation}\label{PsiB}
\Psi_{B} = \nabla^2 \phi - \dfrac{1}{\epsilon^2} (\phi^2 - 1) (\phi +  \sqrt{2}\epsilon m) \; ,
\end{equation}
and
\begin{equation}\label{Gauss}
E_{G}[\phi] = k_{G} \, \dfrac{35}{16 \sqrt{2}} \, \epsilon^3 \, \int_{\Omega} \Psi_{G} \; dV,
\end{equation}
\begin{equation}\label{PsiG}
\begin{split}
\Psi_{G} & = \dfrac{ \bm{\nabla} |\bm{\nabla} \phi|^2 \cdot \bm{\nabla} |\bm{\nabla} \phi|^2}{2} - (\bm{\nabla} |\bm{\nabla} \phi|^2 \cdot \bm{\nabla} \phi) \nabla^2 \phi \\
& + \; |\bm{\nabla} \phi|^2 \bigg[(\nabla^2 \phi)^2 + \bm{\nabla} \phi \cdot \bm{\nabla} \nabla^2 \phi - \dfrac{\nabla^2 |\bm{\nabla} \phi|^2}{2}\bigg] \; .
\end{split}
\end{equation}
Here $m$ is the spontaneous curvature of the membrane, taken to be positive if the membrane bulges towards the 
exterior, $k$ is the bending rigidity and $k_G$ is the Gaussian curvature modulus. Henceforth, we will assume 
$k_G = -k$. $E_B[\phi]$ was already introduced in \cite{Du2004APF} to model the bending energy of the membrane while $E_G[\phi]$ is the new term proposed here to account for the Gaussian energy.

As anticipated, our purpose here is to show that, under the general ansatz (\ref{ansatz}) and in the sharp-interface limit ($\epsilon / D_{ve} = \lambda << 1$), minimizing the phase-field free energy functional (\ref{Energy}) is equivalent to minimizing the Canham-Helfrich free energy. Denoting with a prime the derivative done with respect to $d^*(\bm{x})$, a direct computation leads to
\begin{equation}
\begin{split}
E_{B}[\phi] = \; & k \, \dfrac{3}{4 \sqrt{2}} \, \lambda \, \int_{\bar{\Omega}} \bigg[ \dfrac{1}{\lambda^2}\bigg(f'' - (f^2-1)f\bigg) \; + \\ &+ \; \dfrac{1}{\lambda}\bigg(f' \, \bm{\bar{\nabla}} \cdot \bm{n} + (1-f^2)\sqrt{2}\bar{m}\bigg) \bigg]^2 \, d\bar{V},
\end{split}
\end{equation}
\begin{equation}
E_{G}[\phi] = k_G \, \dfrac{35}{16 \sqrt{2}} \, \int_{\bar{\Omega}} \dfrac{f'^4}{\lambda} \bigg[ (\bm{\bar{\nabla}} \cdot \bm{n})^2 + \bm{n} \cdot \bm{\bar{\nabla}} (\bm{\bar{\nabla}} \cdot \bm{n}) \bigg]  \, d\bar{V},
\end{equation}
where we have denoted with a bar the dimensionless lengths obtained by dividing by $D_{ve}$.
Therefore, in order to minimize $E = E_{B} + E_{G}$, as $\lambda \to 0$, the leading-order term $f_{0}$ of $\phi(\bm{x}) = f(d^*(\bm{x})) = f_{0}(d^*(\bm{x})) + \sum_{i=1}^{+ \infty} \lambda^{i} f_{i}(d^*(\bm{x}))$ must satisfy
\begin{equation}\label{eq_tanh}
f_{0}'' = (f_{0}^2 - 1)f_{0},
\end{equation}
which has the solution
\begin{equation}\label{f0}
f_{0}(d^*(\bm{x})) = \tanh\bigg(\dfrac{d(\bm{x})}{\epsilon \sqrt{2}}\bigg).
\end{equation}
Hence, $\epsilon$ is actually related to the width of the interface. Moreover, by repeating the computations done in \cite{Wang2008AsymptoticAO} for the bending energy alone, it is possible to show that, also in the presence of the new Gaussian energy term, one finds $f_1(d^*(\bm{x})) \equiv 0$ (see the Supplemental Material \cite{Supplemental_Material} for the whole computation). Therefore, given that 
\begin{equation}
\sqrt{2} f_{0}' = (1 - f_{0}^2),
\end{equation}
we are left with
\begin{equation}\label{EBf}
E_{B}[\phi] = k \dfrac{3}{4 \sqrt{2}} \int_{\bar{\Omega}} \dfrac{f_{0}'^2}{\lambda} (\bm{\bar{\nabla}} \cdot \bm{n} + 2\bar{m})^2 \, d\bar{V} \, + \, O(\lambda),
\end{equation}
\begin{equation}\label{EGf}
\begin{split}
&E_{G}[\phi] = \\
&k_G \dfrac{35}{16 \sqrt{2}} \int_{\bar{\Omega}} \dfrac{f_{0}'^4}{\lambda} \bigg[ (\bm{\bar{\nabla}} \cdot \bm{n})^2 + \bm{n} \cdot \bm{\bar{\nabla}} (\bm{\bar{\nabla}} \cdot \bm{n}) \bigg] d\bar{V} \, + \, O(\lambda^2).
\end{split}
\end{equation}
Denoting with $k_1$ and $k_2$ the principal curvatures, we have $\bm{\nabla} \cdot \bm{n} = -(k_1 + k_2) = -2M$ and $\bm{n} \cdot \bm{\nabla} k_i = k_i^2$, with the result that $(\bm{\nabla} \cdot \bm{n})^2 + \bm{n} \cdot \bm{\nabla} (\bm{\nabla} \cdot \bm{n}) = 2 k_1 k_2 = 2G$. Now, noticing that for $\lambda \to 0$
\begin{equation}\label{delta}
\dfrac{f_{0}'^2(\bar{d}(\bm{x})/\lambda)}{\lambda} \stackrel{\mathcal{W}}{\longrightarrow} \; \dfrac{2 \sqrt{2}}{3} \, \delta(\bar{d}(\bm{x})),
\end{equation}
\begin{equation}\label{delta2}
\dfrac{f_{0}'^4(\bar{d}(\bm{x})/\lambda)}{\lambda} \stackrel{\mathcal{W}}{\longrightarrow} \; \dfrac{8 \sqrt{2}}{35} \, \delta(\bar{d}(\bm{x})),
\end{equation}
note \footnote{In Eqs.~(\ref{delta}),~(\ref{delta2}), $\delta(x)$ is the Dirac delta function and $\mathcal{W}$ denotes a weak limit in the sense of distributions.}, and getting back to dimensional variables, the asymptotic behavior follows as
\begin{equation}
E[\phi] \; \sim \; 2k \int_{\Gamma} (M - m)^2 \, dS \; + \; k_G  \int_{\Gamma} G \, dS,
\end{equation}
i.e., the phase-field energy functional reproduces the Canham-Helfrich free energy in the sharp-interface limit ($\epsilon / D_{ve} << 1$). It is worth noticing that the inclusion of the Gaussian energy, which is subdominant in $\lambda$, preserves the hyperbolic tangent form (\ref{f0}) of the leading order solution together with $f_1(d^*(\bm{x})) \equiv 0$, as for the more standard model  with the bending energy alone \cite{Du2004APF}.
Since $f_1(d^*(\bm{x})) \equiv 0$, the desired expression of the bending energy is retained at order $\lambda^{-1}$, and the accuracies $O(\lambda)$ and $O(\lambda^2)$ are guaranteed in Eqs.~(\ref{EBf}),~(\ref{EGf}), respectively. Furthermore, in our formulation, the phase-field Gaussian energy (\ref{Gauss}) has no singularities and actually depends at most on derivatives of order two, as it is possible to see by replacing $\bm{\nabla} \phi \cdot \bm{\nabla} \nabla^2 \phi$ with  $\nabla^2 |\bm{\nabla} \phi|^2/2 - \bm{\mathrm{H}}_\phi:\bm{\mathrm{H}}_\phi$ in (\ref{PsiG}), where $\bm{\mathrm{H}}_\phi$ is the Hessian matrix of the field.

In many cases, it is important to fix the vesicle area and volume. Indeed, since lipids are insoluble in water, the number of membrane lipids is conserved. This fact, coupled with the observation that the membrane rupture tension is very small, implies that the vesicle area $A$ cannot substantially vary at a fixed temperature. 
The volume $V$ of the vesicle is instead determined by the osmotic conditions. 
 In order to enforce the above constraints, in the phase-field context, one can use the functionals
\begin{equation}\label{area}
A[\phi] = \dfrac{3}{4 \sqrt{2}} \, \epsilon \, \int_{\Omega} \bigg[ \dfrac{(1 - \phi^2)^2}{2\epsilon^2} \, + \, |\bm{\nabla} \phi|^2 \bigg] \, dV,
\end{equation}
\begin{equation}\label{volume}
V[\phi] = \int_{\Omega} \dfrac{(1 + \phi)}{2} \, dV,
\end{equation}
which respectively behave like the vesicle area and volume in the sharp-interface limit.

Throughout the paper, an asterisk will denote the dimensionless quantities obtained using $\epsilon$ as the reference length and $8 \pi k$ as the reference energy. The latter is the bending energy of an isolated sphere. A typical value of the bending rigidity is $k = 20 \, k_B T$, with $k_B$ the Boltzmann constant and $T$ the temperature.

\begin{figure}[H]
	\includegraphics[width=8.6cm]{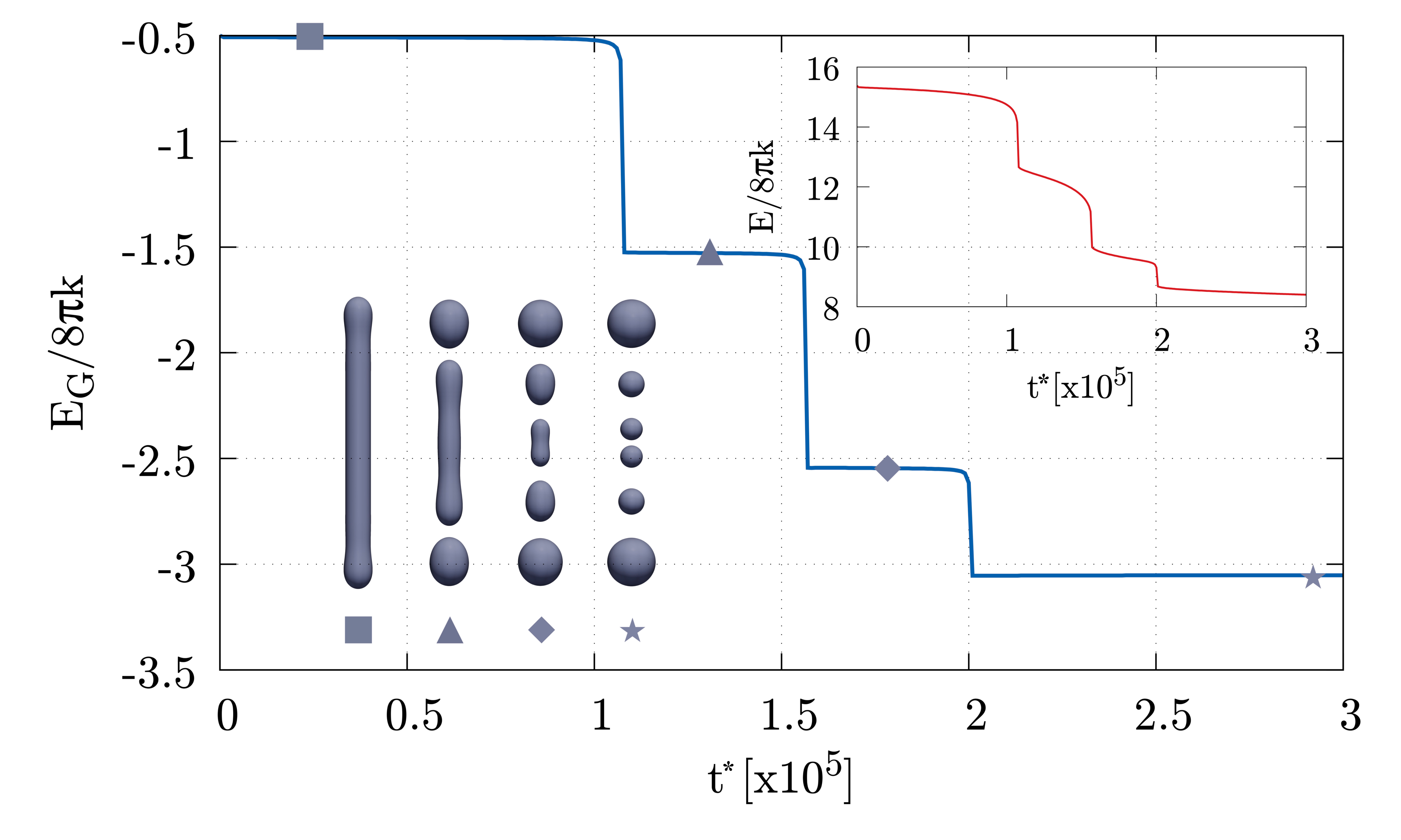}
	\caption{The phase-field Gaussian energy during a series of scissions of a prolate shape into several spheres. The energy jumps by $-4 \pi k$ for any division as prescribed by the Gauss-Bonnet theorem ($k = - k_G$). The fission process occurs due to the presence of a spontaneous curvature $m^* \approx 0.42$. Time evolution is given by the Allen-Cahn gradient flow with $M^* = 8$ (see Section \ref{Numerical_scheme} for more details on the dynamics, the adopted numerical scheme and dimensionless quantities). The inset shows the total energy $E = E_{B} + E_{G}$, which monotonically decreases in time, revealing the stability of the scheme. This z-axial symmetric simulation has been carried out in a $[0, \, 36]\times[0, \, 440]$ computational domain in the $r^*-z^*$ plane with a $54\times660$ mesh, initial $D_{ve}^* = 1/\lambda \approx 109$ and $dt^* = 4$. There is no constraint on the area, which, at the end of the simulation, differs from the initial value by approximately $6.9\%$. Volume is conserved with a relative error smaller than $10^{-7}$ with respect to its initial value.}
	\label{gaussdim}
\end{figure}

Figure~\ref{gaussdim} shows the Gaussian energy during a series of scissions of an unstable prolate shape into several spheres. The process is driven by the spontaneous curvature. The evolution equation is described in Section~\ref{Numerical_scheme} together with the adopted numerical scheme. Details concerning the numerical validation are provided in Appendix~\ref{Numerical_validation}. Consistency of the present phase-field approach with the Gauss-Bonnet theorem is discussed in Appendix~\ref{Gauss_Bonnet_appendix}. Here, it is only worth saying that the novel model is able to properly capture the Gaussian energy jumps due to topological transitions. 

\begin{figure*}
	\includegraphics[width=\textwidth]{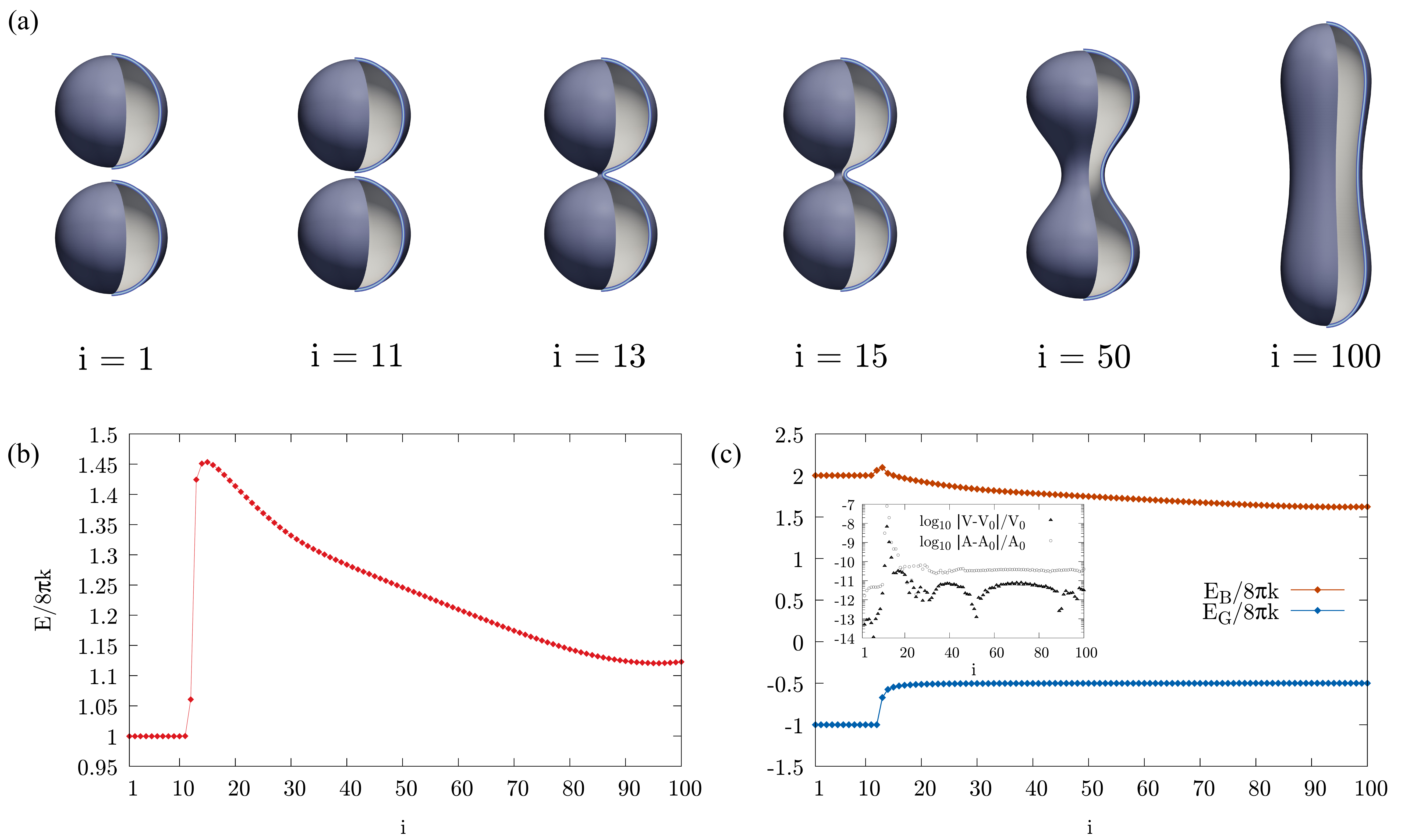}
	\caption{The minimal energy pathway connecting two spheres of radius $R^* = 87.5$ with a dumbbell shape, $k = - k_G$. The path consists of vesicles with constant area and volume and therefore with constant reduced volume $v \approx 1/\sqrt{2}$. There is no spontaneous curvature, $m^* = 0$. This z-axial symmetric result is obtained with the string method and the new phase-field model using a $[0, \, 96] \times [-245, \, 245]$ computational domain in the $r^*-z^*$ plane with a grid of $144\times735$ nodes per image, $N = 100$ images and $1/\lambda \approx 247.5$. (a) Six vesicle shapes along the minimal energy path, identified by their image number $i = (N-1)\alpha_i + 1$, being $\alpha$ the string parameter (equal arc-length parameterization). From right to left we can observe the fission process of the dumbbell shape into two spheres, whereas from left to right the fusion process. (b) The phase-field energy, Eq. (\ref{Energy}), along the path. Saddle point is placed between the images $i = 14$ and $i = 15$ and consists of two spherical vesicles connected by a catenoid-like neck. (c) The bending and Gaussian energy contributions to the energy along the path. The inset shows the effectiveness of the scheme in preserving vesicles area and volume. Reference values of area and volume are $A_0^* = 1.924392 \cdot 10^5$ and $V_0^* = 5.615982 \cdot 10^6$.
	}
	\label{fig:stringpanel}
\end{figure*}

\section{Results}\label{Results}

In the topological transition between two spherical vesicles and a dumbbell-shaped one, which are two stable states, the system goes through a sequence of configurations $\phi_\alpha({\bm{x}})$ in the space of the phase-field, identifying a path which we parameterize by the normalized arc-length $\alpha \in [0, 1]$. An MEP for this transition is a curve on the energy landscape $E[\phi]$ connecting the two stable states $\phi_{\alpha=0}(\bm{x})$ and $\phi_{\alpha=1}(\bm{x})$, respectively, and such that it is everywhere tangent to the gradient of the potential ($\partial \phi_\alpha/\partial \alpha \propto \delta E[\phi_\alpha]/\delta \phi$), except at critical points \cite{Cameron2011TheSM}. An initial guess of the path is discretized in a \textit{string} made up of $N = 100$ images corresponding to $\alpha_i = (i-1)/(N-1)$. The initial guess is relaxed towards the MEP by means of the string method (see \cite{Weinan2002StringMF, maragliano2006string, E2007SimplifiedAI} and Section \ref{String_method}) suitably accounting for the constraints of constant total surface area, Eq. (\ref{area}), and enclosed volume, Eq.~(\ref{volume}). The obtained MEP goes through a saddle point $\phi_{\alpha_c}(\bm{x})$ for the free energy, determining the transition barriers $\Delta E^\dagger_{0 \to 1} = E[\phi_{\alpha_c}] - E[\phi_{\alpha=0}]$ and $\Delta E^\dagger_{1 \to 0} = E[\phi_{\alpha_c}] - E[\phi_{\alpha=1}]$, for the forward and backward process, respectively.

\begin{figure*}[t!]
	\includegraphics[width=\textwidth]{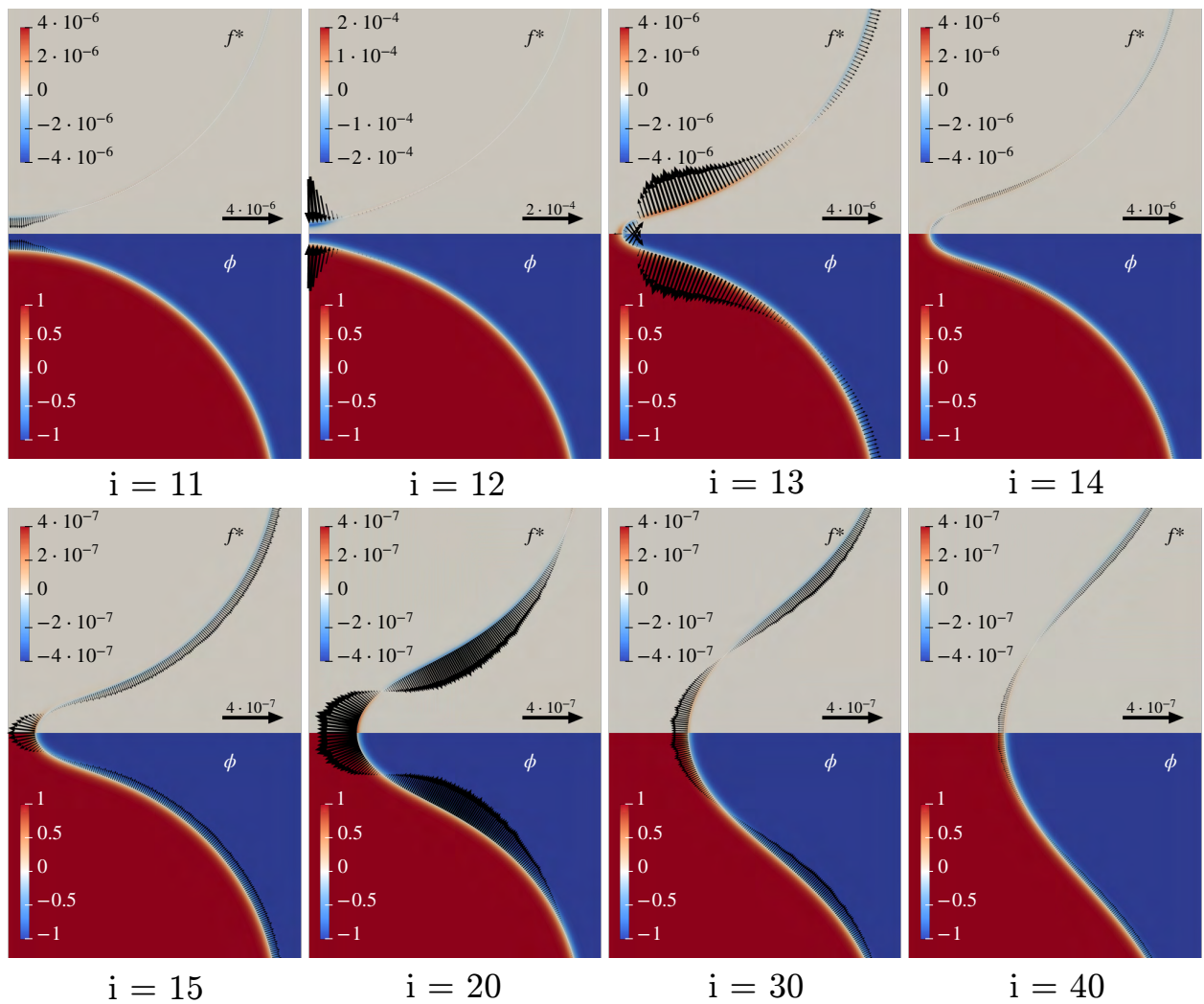}
	\caption{Detailed views in the $r^*-z^*$ plane of the vesicle configurations along the MEP. The index $i = (N-1)\alpha_i + 1$ numbers the images on the string. Vectors, plotted for clarity only on the $\phi = 0$ isoline, provide the force field $\bm{f} = -\delta E/\delta \phi \bm{\nabla} \phi$  required to keep the vesicle in equilibrium in the given configuration, balancing the internal elastic reaction. The contours in the upper part of each view show the normal component of the force, while those on the lower part depict the field $\phi$. For a better visibility, vectors are scaled according to the reference arrow in each plot.
	}
	\label{fig:forces}
\end{figure*}
Figure~\ref{fig:stringpanel} shows the computed MEP for membranes with zero spontaneous curvature, $m = 0$. Since the phase-field $\phi$ reaches its limiting values $\pm 1$ with an accuracy of about $3 \%$ already at a distance of $\pm 3 \epsilon$ from the $\phi = 0$ membrane midsurface, we assume that $l_{pf} = 6 \epsilon$ represents the thickness of the diffuse interface. We have shown in Section \ref{PFMD} that the phase-field description recovers the Canham-Helfrich model in the limit of small $\lambda \propto \ell_{pf}/D_{ve}$. Our numerical experiments, Appendix \ref{Numerical_validation}, point out that this asymptotic behavior is already achieved when $\ell_{pf}/D_{ve} >> (\ell_{me}/D_{ve})_{\max} = 1/40$, the latter being the maximum thickness-to-curvature radius ratio for which the Canham-Helfrich model is accepted \cite{Dimova2019TheGV}. Since the relative distance between approaching membrane segments is relevant during the topological transition, it is crucial that the diffuse interface width matches the bilayer thickness. This requirement fixes the scale of our system. Setting $l_{pf} = l_{me} = 5 \, \text{nm}$, the configurations shown in Figure \ref{fig:stringpanel}a correspond to vesicles with $D_{ve} \approx 206 \, \text{nm}$, thus within the range of validity of the asymptotic Canham-Helfrich model.

Fig.~\ref{fig:stringpanel}a shows successive configurations along the MEP. Increasing/decreasing $\alpha$ corresponds to moving along the path in the direction of the fusion/fission (forward/backward) process, respectively. Proceeding forward, the two vesicles come closer to each other without deforming, get in touch, and merge together forming a narrow neck that expands until the final dumbbell-shaped configuration is reached. As explained in Section \ref{INTRO}, the equilibrium states of a vesicle are determined by its reduced volume and reduced spontaneous curvature, which, in the present case, are $v = 1/\sqrt{2}$ and $m_0 = 0$, respectively, where $1/\sqrt{2}$ is the only reduced volume compatible with a vesicle obtained from the fusion of two spheres of the same radius. As shown in \cite{Seifert1991ShapeTO}, with these parameters, it is possible to reach two axisymmetric configurations with the topology of a sphere, namely one oblate-discocyte shape and one prolate-dumbbell shape. The latter has the lowest energy and, in the present case, is the equilibrium state assigned to the string as the final configuration, $\phi_{\alpha = 1}(\bm{x})$.

Figure~\ref{fig:stringpanel}b shows the free energy profile along the MEP. The free energy of the final configuration (prolate) is $E[\phi_{\alpha = 1}]/(8 \pi k) \approx 1.12$, which is larger than the initial energy $E[\phi_{\alpha=0}]/(8 \pi k) = 1$ of the two spheres. Both values are in excellent agreement with the data reported in \cite{Seifert1991ShapeTO}. One may notice that the two-spheres configuration possesses a sequence of neutral equilibrium states, corresponding to rigid translations during which the two vesicles approach/separate from each other (configurations $i$ from $1$ to $11$, as also depicted in Fig.~\ref{fig:stringpanel}a). The saddle point consists of two spheres connected by a small narrow neck and is located between configurations $i = 14$ and $i = 15$, with the latter having the highest energy of the two, $E[\phi_{\alpha = \alpha_c}]/(8 \pi k) \approx 1.45$. It should be noticed that such a configuration possesses the bending energy of two spheres together with the Gaussian energy and the topology of a single sphere. Hence, the forward and backward computed free energy barriers are $\Delta E_{0 \to 1}^\dagger/(8 \pi k) \approx 0.45$ and $\Delta E_{1 \to 0}^\dagger/(8 \pi k) \approx 0.33$, respectively. Considering a bending rigidity $k$ of order $20 \,  k_B T$ \cite{Dimova2014RecentDI}, it turns out that both fusion and fission processes require further agents in order to happen, in addition to elasticity and thermal fluctuations. These agents are typically protein systems. Still in Fig.~\ref{fig:stringpanel}b, it is possible to observe a substantial asymmetry between fusion and fission, with a much steeper energy increase required to reach the transition state in the fusion process.

The main plots in Fig.~\ref{fig:stringpanel}c provide the bending and Gaussian contributions to the free energy along the MEP. Apparently, the forward barrier $\Delta E_{0 \to 1}^\dagger$ is almost entirely due to the Gaussian energy jump associated with the topological change. On the other hand, the backward barrier $\Delta E_{1 \to 0}^\dagger$ builds up continuously with the progressive deformation of the prolate shape to form the narrow neck preceding the actual fission. The inset shows the evolution of the area and enclosed volume along the MEP, confirming that the constraints are perfectly satisfied at each string image. 

Figure~\ref{fig:forces} focuses on the region of the MEP where the most relevant events associated with the topological transition take place, images $i = 11, \, \ldots \, , \, 40$. The contour plots in the lower half panels of Fig.~\ref{fig:forces} provide the structure of the phase-field as a function of radius $r^*$ and axial coordinate $z^*$, with $\phi$ smoothly joining the inner region $\phi = 1$ to the outer region $\phi = -1$ through the layer of dimensionless thickness $\ell_{pf}^* = 6$.

As explained in Section~\ref{Force_Field_Computation}, each image of the string can be rendered a state of equilibrium by introducing a force field $\bm{f} = -\delta E/\delta \phi \bm{\nabla} \phi$ that counterbalances the membrane elastic reaction. Considering the forward transition, $0 \to 1$,  such force field from $\alpha = 0$ to $\alpha = \alpha_c$ can be interpreted as the external force needed to drive the transition under quasi-static conditions, thus spending the minimal work ${\cal W}_{0 \to 1} = \Delta E^\dagger_{0 \to 1}$. Once the critical state is overcome, the system can be left to evolve spontaneously until it reaches the final equilibrium state $\alpha = 1$. Symmetric considerations hold for the backward transition $1 \to 0$. The dimensionless vector fields $\bm f_\alpha^*({\bm x})$ are depicted as arrows in each panel of Fig.~\ref{fig:forces}, where, for the sake of better readability, they are  plotted only on the $\phi = 0$ isoline. The contour plots on the upper part of each  panel display the component of the force normal to $\phi$-isolines. It should be noticed that the scale of the arrows changes from panel to panel, at least for the upper frames, $i = 11, \, \ldots \, , \, 14$. For the forward process, the latter are the configurations achieved just before the critical state. In this region, the MEP is particularly steep, requiring more intense forces, which result to be strongly localized near the vesicles contact region. On the contrary, the backward process requires a more distributed force field, as shown in images $i = 15, \, \ldots \, , \, 40$. The arrows reverse their direction between configurations $i = 14$ and $i = 15$, showing that in this interval the force field vanishes, confirming that the critical state occurs somewhere between these two images.

\section{Discussion}\label{Discussion}

As shown in Fig.~\ref{fig:stringpanel}, the phase-field model is able to account for the Gaussian energy and its effects on the fusion and fission processes, allowing to bridge the gap across the topological transition. The initial and final (meta)stable states of the process under consideration, i.e. the two spheres and the dumbbell-shaped vesicle, are perfectly consistent with the predictions of the sharp interface description {\sl à la} Canham-Helfrich \cite{Seifert1991ShapeTO}, both in terms of energy and shape. The two states have the same reduced volume $v = 6 \sqrt{\pi} V/A^{3/2} = 1/\sqrt{2}$, which is constant throughout the process, since both the volume and the area are accurately conserved, inset of Fig.~\ref{fig:stringpanel}c.

The force field $\bm f = - \delta E/\delta \phi \bm \nabla \phi $ is able to induce both the fusion and fission processes with a minimal work expenditure. This external driving force, depicted in Fig.~\ref{fig:forces}, is clearly required only during the climbing phases of the energy landscape, Fig.~\ref{fig:stringpanel}b, i.e. from $i = 1$ (two spheres configuration) to $i = 15$ (saddle point) for fusion and from $i=100$ (dumbbell-shaped vesicle) back to $i = 15$ for fission. Proceeding in the fusion direction, the energy landscape is initially flat and concerns the apposition of the two vesicles. In the biological context, the final part of this stage corresponds to the dehydration of the two facing membranes. This effect is clearly absent in our model, although it could be included by introducing a suitable effective repulsive potential. After this flat region, the energy undergoes a sharp jump, which is related to the variation of the Gaussian energy across the topological transition, $|\Delta E_G|/(8 \pi k) = 0.5$. This jump is associated with the change of the Euler characteristic $\chi$, with $\chi = 4$ for the two spheres and $\chi = 2$ for the prolate-dumbbell shape, with the latter homeomorphic to a single sphere. Therefore, the force field required to complete the fusion process is very intense and localized, Fig.~\ref{fig:forces}, and essentially due to the Gaussian energy contribution. This leads to an asymmetry between the fusion and fission processes. Indeed, in the former case, the activation energy is entirely due to the topological transition, while in the latter case the barrier progressively builds up through the continuous membrane deformation.

The forces required for overcoming the fusion topological barrier are too strong to be directly exerted by the sole mechanical action of proteins. For example, setting $k = 20 \, k_B T$ \cite{Dimova2014RecentDI}, the resulting activation energy is $\Delta E_{0 \to 1}^\dagger \approx 226 \, k_B T$. Consistently with the present findings, Deserno \cite{bassereau20182018} suggests that fusion proteins, besides a mechanical action, may contribute to lowering the energy barrier by locally modifying the Gaussian modulus in the contact region of the approaching membranes. Indeed, the introduction of a suitable, spatially dependent Gaussian modulus is expected to reduce the stiffness associated with the GB theorem, opening alternative routes to the topological change. Our results show that this scenario is actually possible since the forces associated with the Gaussian energy are localized in the region of contact between the two spheres and, therefore, it is reasonable that a variation of $k_G$ in such a region could lower the activation energy. For example, this situation is compatible with the observation that influenza virus hemagglutinin proteins, in addition to having an apposition activity, are also able to perturb the membrane lipid bilayer by insertion of their amphipathic fusion peptide \cite{tareste2018common}. Interestingly, the present phase-field approach can be easily adapted to the instance of a topological transition with a spatially dependent Gaussian modulus, a case we leave for future work.

\begin{figure}[H]
	\includegraphics[width=8.6cm]{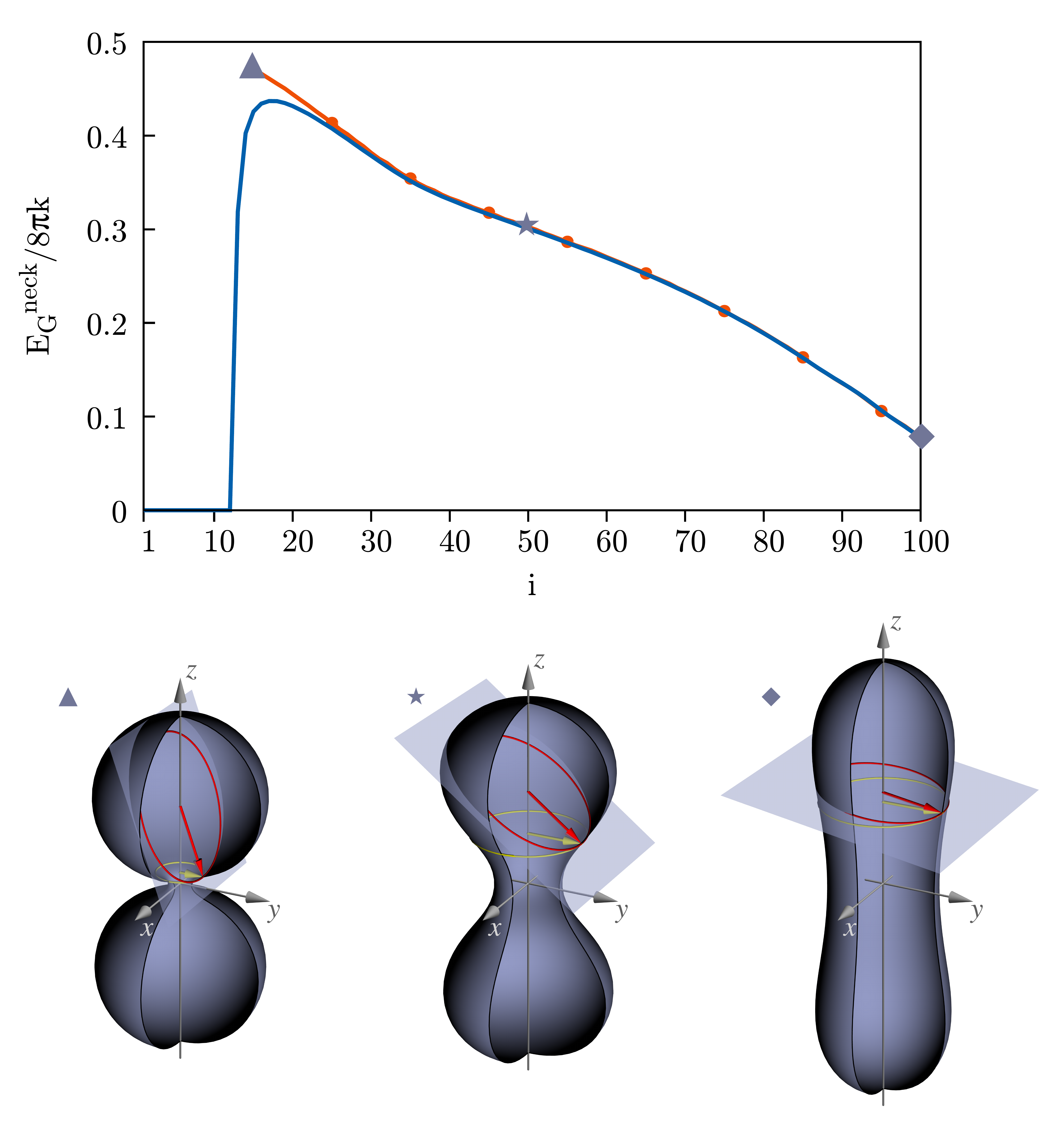}
	\caption{{\small Top panel: normalized Gaussian energy of the neck, Eq.~\eqref{neck_energy}, along the MEP (blue line). The orange line with dots provides the neck Gaussian energy as a post-processing based on the sharp interface Canham-Helfrich energy, Eq.~\ref{Canham-Helfrich}, computed considering the $\phi = 0$ isoline as the membrane midsurface: $E^{\text{neck}}_G(Z) / (8 \pi k) \approx \sqrt{1 - (r(Z)/R_n)^2}/2$. The agreement between the two curves progressively deteriorates when getting closer to the saddle point, due to the increasing curvature of the membrane generatrix. Bottom panel: three membrane configurations sketching the upper (yellow) circular boundary of the neck with its curvature radius $r(Z)$ (yellow arrow), and the osculating (red) circle to the vesicle cross section with the cutting plane passing through the neck boundary and containing both the surface normal and the tangent to the circle. The radius $R_n$ of the osculating circle is shown as a red arrow. The position of each configuration along the MEP is denoted by the corresponding symbol (triangle, star and rhombus).}}
	\label{Neck_energy}
\end{figure}

As illustrated in Fig.~\ref{fig:stringpanel}a, the topological transition is mediated by the formation of a catenoid-like neck \cite{chabanon2018gaussian}, similarly to what has been observed in the experiments \cite{Avinoam2015EndocyticSM}. Operationally, we define the neck region as the z-chunk of the fused vesicle where the local contribution to the Gaussian energy
\begin{equation}\label{neck_energy}
	\begin{split}
	E^{\text{neck}}_{G}(Z) & = \, k_G \, \dfrac{35}{16 \sqrt{2}} \, \epsilon^3 \, \int_{-Z}^{+Z} dz \int 2 \pi r \, \psi_G \, dr \approx \\ & \approx k_G \int_{-Z}^{+Z} G \; j(z) \, dz
	\end{split}
\end{equation}
is positive, $k_G \, G \, j(z) \approx k_G \, \int 2 \pi r \, \psi_G \, dr > 0$. The approximate equality follows by considering that the phase-field functional $E^{\text{neck}}_{G}$ approaches the corresponding sharp-interface Canham-Helfrich Gaussian energy, where the neck midsurface is described by the Monge representation $r = r(z)$ and $dA = j(z) dz$ is the corresponding area element. The computed phase-field Gaussian energy of the neck along the MEP is shown in the top panel of Fig.~\ref{Neck_energy}, blue line. Proceeding from left to right, $E^{\text{neck}}_{G}(Z)/(8 \pi k)$ sharply increases to a value close to (though smaller than) $0.5$ and subsequently decreases. It is worth stressing that, from a sharp interface point of view, the total curvature, namely the integral of the Gaussian curvature, is $4 \pi$ for a single sharp sphere, which corresponds to $E^\text{CH}_G/(8 \pi k) = -0.5$ in terms of Canham-Helfrich Gaussian energy. Given two initially disjoint sharp spheres (total curvature $8 \pi$, $E^{\text{CH}}_G/(8 \pi k) = -1$), the formation of a joining neck changes the topology and reduces the total curvature to that of a single sphere, $4 \pi$, $E^{\text{CH}}_G/(8 \pi k) = -0.5$. There are two main reasons why the phase-field model provides a $E^{\text{neck}}_{G}(Z)/(8 \pi k)$ contribution to the Gaussian energy that is slightly smaller than $0.5$: i) close to the transition state, the curvature of the neck generatrix in the $r-z$ plane is comparable with the finite thickness of the bilayer, so that the sharp-interface model is inappropriate; ii) the total curvature of the neck midsurface is always larger than $- 4 \pi$ and can reach the latter limiting value only when the tangent to the generatrix gets orthogonal to the z-axis at the two edges, see note \footnote{Given the parametric representation ${\bf x} = {\bf x}(\xi^1,\xi^2)$ of a surface ${\cal M}$, the Gaussian curvature is $G \, j = {\bf n} \cdot \partial \bf {n} / \partial \xi^1 \times \partial \bf{n}/\partial \xi^2$, where $j$ is the Jacobian and ${\bf n}$ the unit normal. Introducing the Gauss map, that associates the point ${\bf x} = {\bf x}(\xi^1,\xi^2) \in {\cal M}$ to a corresponding point ${\bf n}(\xi^1,\xi^2)$ on the unit sphere ${\cal S}$, one finds $\int_{\cal M} G dS = \int_{{\cal S}({\cal M})} d\Omega$, where $\Omega$ is the (signed) solid angle and ${\cal S}({\cal M}) $ is the image of ${\cal M}$ through the map. In general, the image of the neck does not completely cover the sphere, since the normal ${\bf n}$ to the neck surface $\cal M$ at the boundary $\partial {\cal M}$ does not get aligned with the z-axis. As a consequence, $\int_{\cal M} G dS \le 4 \pi$.}.

Evidently, the neck contribution to the Gaussian energy is also the main contribution to the barrier $\Delta E_{0 \to 1}^\dagger/(8 \pi k) \approx 0.45$. Proceeding to the right along the MEP, beyond the saddle point, the computed Gaussian energy of the neck progressively decreases, top panel of Fig.~\ref{Neck_energy}, blue line. Since, Fig.~\ref{fig:stringpanel}c, in that region the total Gaussian energy remains overall constant, $E_G/(8 \pi k) = -0.5$, the energy lost by the neck is redistributed to the remaining, dome-like parts of the vesicle. 

Figure~\ref{Neck_energy}, top panel, orange line with dots, also provides the neck Gaussian energy as a post-processing based on the sharp interface Canham-Helfrich energy, Eq.~\ref{Canham-Helfrich}, computed considering the $\phi = 0$ level set as the membrane midsurface. In such a manner, the neck Gaussian energy can be evaluated as
\begin{equation}\label{Sharp_neck_energy}
	\dfrac{E^{\text{neck}}_G(Z)}{8 \pi k} \approx \dfrac{\sqrt{1 - (r(Z)/R_n)^2}}{2} \; ,
\end{equation}
where $r(\pm Z)$ is the radius of the circular neck boundary and $R_n$ is the radius of normal curvature of the vesicle cross section with the plane containing both the surface normal and the tangent to the circles, evaluated at the neck boundary \cite{do2016differential}. In order to get a geometrical understanding of these quantities, the bottom panel of Fig.~\ref{Neck_energy} shows a few vesicle configurations, with their upper neck boundary (yellow circle) and the osculating (red) circle defined by the radius of normal curvature. The yellow and red arrays depict $r(\pm Z)$ and $R_n$, respectively. A more detailed discussion about Eq.~\eqref{Sharp_neck_energy} is provided in the Supplemental material \cite{Supplemental_Material}. The agreement between the phase-field Gaussian energy of the neck and its sharp interface counterpart is excellent as long as different membrane segments do not approach each other to a distance comparable to the bilayer thickness. Indeed, the agreement progressively deteriorates when getting closer to the saddle point, due to the increased curvature of the membrane generatrix. Actually, also in this stage, one could better and better reproduce the sharp interface energy by reducing the regularizing parameter $\lambda$. On the other hand, from a physical point of view, the thickness of the bilayer is finite, making the sharp interface model inappropriate when the saddle point is approached. Overall, these results confirm the accuracy of the proposed phase-field expression for the Gaussian curvature.

\begin{figure}[H]
	\includegraphics[width=8.6cm]{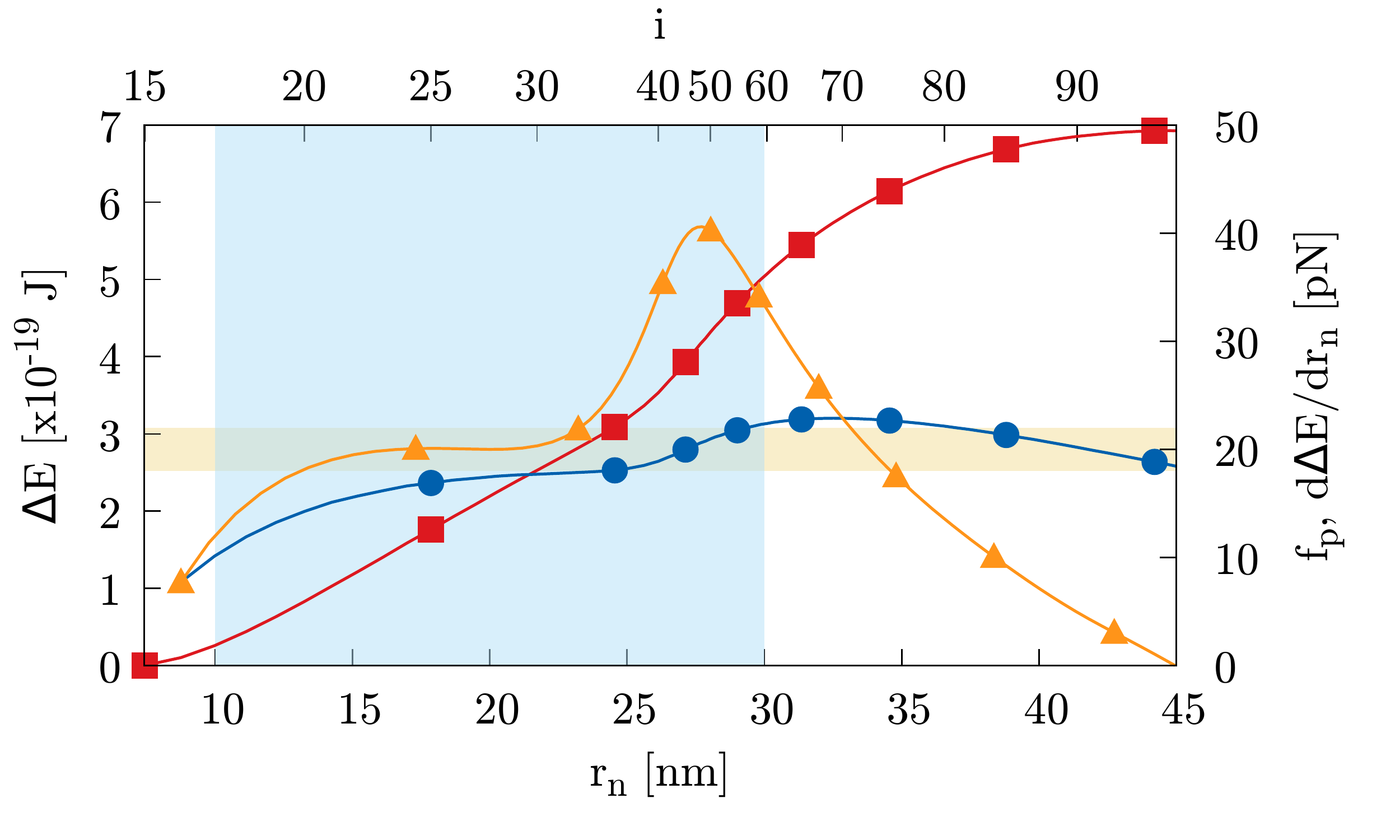}
	\caption{{\small Red curve with squares: energy needed to complete the fission as a function of the current neck radius $r_n$, $\Delta E = E(r_n) - E(r_0)$ vs $r_n$. The second abscissa axis on top of the frame provides the image number $i$ along the MEP. Orange curve with triangles: estimated constriction force (second ordinate axis on the right), $d \Delta E/d r_n$ vs $r_n$. Blue curve with dots: $f_p = \Delta E(r_n)/(r_n -r_0)$ vs $r_n$. The vertical light blue band represents the range in which dynamin polymerizes \cite{roux2010membrane}. The horizontal light orange strip depicts the value of dynamin constriction force, measured in the experiments \cite{roux2010membrane, Steinkhler2020ControlledDO}.
	}}
	\label{force_protein}
\end{figure}

As anticipated, the forces at play during fission are more distributed and less intense than for fusion. The large region they act on,  Fig.~\ref{fig:forces}, is consistent with the cooperation of several protein systems, like, e.g., in clathrin mediated endocytosis, which involves clathrins polymerization and the subsequent action of the constrictase dynamin \cite{Avinoam2015EndocyticSM}. One can estimate the minimal work the protein system needs to perform to induce the topological change by comparing the free energy barrier $\Delta E_{1 \to 0}^\dagger$ with the protein work $\mathcal{W}_{1 \to 0} = f_{p} \, \Delta r$, where $f_{p}$ is the order of magnitude of the protein force and $\Delta r = r_{max} - r_0$ is the change in vesicle radius at the neck, between the equilibrium prolate ($r_{max}$) and the saddle point configurations ($r_0$). Given the scale of the system, Section~\ref{Results},  we find
$\Delta r = 37.4 \, \text{nm}$ which, from the barrier height, provides $f_p = 0.91 \, k \, \text{pN}/k_BT$. Interestingly, for the values of $k$ proper of fluid lipid membranes, we obtain protein forces in fairly good agreement with the experimental estimates reported in \cite{Steinkhler2020ControlledDO}, e.g. $ \simeq 20 \, \text{pN}$ for dynamin, $\simeq 65 \, \text{pN}$ for ESCRT-III and $\simeq 80 \, \text{pN}$ for FtsZ. For  example, by assuming $k = 20 \, k_B T$, we obtain a protein constriction force $f_{p}$ of $18.2 \, \text{pN}$. For the same bending rigidity, Fig. \ref{force_protein} shows, red curve with squares, the energy needed to complete the fission process as a function of the current neck radius $r_n$, $\Delta E(r_n) = E(r_n) - E(r_0)$ (note that the fission proceeds from larger to smaller neck radii,  i.e. from right to left along the abscissa). The corresponding image number $i$ along the MEP is provided on the second abscissa axis on the top of the frame. The slope of the plot, $d\Delta E/d r_n$, orange line with triangles, provides the estimate of the constriction force (positive when constrictive). A plateau is apparent at $d \Delta E/d r_n \simeq 20 \, \text{pN}$ in the range of radii $16 \le r_n \le 21 \, \text{nm}$. Notably, it is known from the literature \cite{roux2010membrane} that, e.g., dynamin polymerizes on tubules with radius between $10$ and $30 \, \text{nm}$, exherting forces of the order of $20 \, \text{pN}$. In order to facilitate comparison with published data, Fig.~\ref{force_protein} also provides in blue, with dots, $f_p = \Delta E(r_n)/(r_n -r_0)$.

It is worth stressing that during topological transitions, when the relative distance between approaching membrane segments becomes comparable to the bilayer thickness, the scale invariance of the asymptotic Canham-Helfrich Hamiltonian (\ref{Canham-Helfrich}) is broken. For such a reason, Section~\ref{Results}, we defined the scale of our system by matching the bilayer thickness with the diffuse interface width. However, far from the topological change, scale invariance is preserved, implying that, during the phases where the Gaussian energy remains constant, the results of Figs.~\ref{fig:stringpanel} and \ref{fig:forces} hold for vesicles larger than those considered so far, too. Furthermore, for vesicles of smaller $\lambda = \epsilon / D_{ve}$, the saddle point energy $E[\phi_c]$ ought to approach the limiting value $1.5$, namely the energy of two spheres connected by a zero radius neck.

So far, we have discussed the fusion and fission processes as obtained by relaxing the system to the MEP using the Allen-Cahn dynamics, endowed with the additional constraints on surface area and enclosed volume. Possible alternatives can be considered, e.g., the Cahn-Hillard dynamics where the relaxation is driven by the divergence of a flux. Both choices lead to the same free energy barriers since the critical configurations (local minima and saddle point) are the same. However, the path connecting the saddle point to the two minima depends on the specific relaxation dynamics and the  Allen-Cahn relaxation does indeed provide the minimal energy path.

\section{Final Remarks}\label{Final Remarks}

Let us conclude with a few general remarks. We have described the full-scale process of topology change in the fusion/fission process of two large unilamellar vesicles (LUVs) with an approach that can be extended to deal with giant unilamellar vesicles (GUVs). The model thoroughly accounts for the Gaussian energy, for which a suitable phase-field expression has been provided, and, far from the topology change, recovers the Canham-Helfrich description in the limit of small bilayer thickness. During the transition, our proposal should be interpreted as a rational way to regularize the singularity, leading to a process that smoothly matches the external solution before and after the transition. The model clearly misses the many molecular details associated with the dynamics of the lipids forming the bilayer. It could be argued, however, that the correction due to such details should be small as compared to the energy barrier associated with the full-scale evolution of the vesicle. 

Naively, one may argue that proteins systems could have evolved to overcome the large barrier that stabilizes the 
vesicle topology by following a minimal energetic pathway. Hence, we have evaluated the minimal free energy path for the transition and extracted the force field able to drive the process with minimal work expenditure. The free energy profile we find confirms the strong asymmetry between the fusion and the fission processes. For fusion, the force field required to overcome the estimated barrier is too intense to be exerted by protein systems and calls for additional effects that could locally modify the Gaussian modulus during the topological change. On the contrary, the spatial scales and the forces acting during fission are consistent with the experimental estimates for typical fission protein systems.

Finally, it can be noted that the diffuse interface approach can naturally be coupled with hydrodynamics \cite{anderson1998diffuse, magaletti2013sharp} to include the dynamic effect of the external and internal aqueous environments of the vesicles. One may also stress that the Gaussian energy functional can find a much broader scope, e.g., as an indicator of the topological genus in the context of cluster analysis \cite{hansen1997cluster, frades2010overview}, or as a way to provide a barrier towards undesired/unphysical fusion processes. A compelling example concerns emulsions where surfactant-covered droplets behave much like lipid micelles \cite{thiam2013biophysics, fuller2012complex}, suggesting that the Gaussian energy is the reason why surfactants do act as emulsifiers.

\section{Methods}\label{Methods}

\subsection{Numerical scheme}\label{Numerical_scheme}

The numerics relies on FFT-based spectral differentiation in cell-centered grids which provide
high accuracy solutions, with special regard to the estimate of the Gaussian energy.
The accuracy in evaluating the Gaussian energy, Eq.~(\ref{Gauss}), is shown in Table~\ref{Table_gauss}
for a sphere, a torus and a straight cylinder. Given the axial-symmetry of these shapes, all the computations are done in a $[0, 40] \times [0, 40]$ computational domain in the $r^*-z^*$ plane with a grid of $80 \times 80$ nodes. 
In evaluating the functional,  we set $\phi(\bm{x^*}) = \tanh((\sqrt{r^{*2} + (z^*-20)^2} -10)/\sqrt{2})$ for the sphere, $\phi(\bm{x^*}) = \tanh((\sqrt{(r^*-20)^2 + (z^*-20)^2}-10)/\sqrt{2})$ for the torus and $\phi(\bm{x^*}) = \tanh((r^*-10)/\sqrt{2})$ for the cylinder, which are obtained using Eq.~(\ref{f0}).

\begin{table}[H]
\centering
\caption{\label{Table_gauss} Gaussian energy computed values, $k_G = -k$.}
\normalsize
\begin{tabular}{p{2.00cm} c c}
	\hline
	\hline
	\rule[-1ex]{0pt}{4ex}
    Shape & $E_{G}/8\pi k$ (exact)  & $E_{G}/8\pi k$ (numerical)\\
    \hline
	\rule[-1ex]{0pt}{4ex}
    Sphere & $-5. \cdot 10^{-1}$ & $-5.000525\cdot10^{-1}$ \\

	\rule[-1ex]{0pt}{4ex}
	Torus & $0.$ & $-1.729446\cdot10^{-18}$ \\

	\rule[-1ex]{0pt}{4ex}
	Cylinder & $0.$ & $-9.860761\cdot10^{-32}$\\
	\hline
	\hline
\end{tabular}
\end{table}

The energy pathways of Section~\ref{Results} are obtained by means of the string method, which is briefly described in Section~\ref{String_method}. The remaining simulations reported in this paper, i.e. the one shown in Fig.~\ref{gaussdim} and those in Appendix~\ref{Numerical_validation}, are carried out using the Allen-Cahn dynamics
\begin{equation}\label{Allen-Cahn}
\dfrac{\partial \phi}{\partial t} = - M \dfrac{\delta \bar{E}}{\delta \phi},
\end{equation}
where $M$ is the mobility coefficient and $\delta \bar{E} / \delta \phi$ is the functional derivative of the augmented energy
\begin{equation}\label{E_modified}
\begin{split}
& \bar{E}[\phi] = E[\phi] \, + \\ & + \, \gamma(A[\phi]- A_0) + \dfrac{1}{2} M_1 (A[\phi] - A_0)^2 \, + \\ & + \, \Delta p (V[\phi] - V_0) + \dfrac{1}{2} M_2 (V[\phi] - V_0)^2 \; .
\end{split}
\end{equation}
Here, the additional terms added to the energy (\ref{Energy}) are needed when constraining to $A_0$ and $V_0$ the vesicle area (\ref{area}) and volume (\ref{volume}), respectively. $M_1$, $M_2$ are two penalty constants, whereas $\gamma$ and $\Delta p$ are updated at each time step according to the \textit{augmented Lagrangian method}, \cite{Constrained_string_method}:
\begin{align}
& \gamma^{n+1}  \; \; \, = \; \gamma^n + M_1 (A[\phi^{n+1}] - A_{0}), \\	
& \Delta p^{n+1}  = \; \Delta p^n + M_2 (V[\phi^{n+1}] - V_{0}).
\end{align}
Therefore $\gamma$ and $\Delta p$ are estimates of the Lagrange multipliers that improve at every time step. Starting 
from an assigned initial condition, the Allen-Cahn dynamics causes the energy to monotonically decrease in time until it 
reaches a critical steady-state. The dimensionless time and mobility are $t^* = t/\tau_R$ and $M^* = 8 \pi k M \tau_R / 
\epsilon^3$, respectively, with $\tau_R$ a suitable time scale.

With the help of the PETSc library \cite{PETSc}, a Crank-Nicolson time-stepping scheme is employed to integrate the Allen-Cahn gradient flow, while a semi-implicit Euler single step scheme is used to solve the more computationally demanding string dynamics. The explicit form of the functional derivative $\delta \bar{E} / \delta \phi$ is given in Appendix~\ref{Appendix_functional_derivatives}.

\subsection{String method}\label{String_method}

The zero-temperature string method \cite{Weinan2002StringMF} is a technique for computing free energy barriers and transition pathways on a given energy landscape. The method proceeds by evolving in time a  \textit{string}, namely a curve parameterized by $\alpha \in [0, 1]$. For each $\alpha$ the image of the string is a phase-field function 
$\phi_\alpha(\bm{x})$ representing a membrane state.

Given an initial guess for the pathway connecting two local minima, the string evolves in time following the dynamics
\begin{equation}
	\dfrac{\partial \phi_{\alpha}}{\partial t} = - M \bigg(\dfrac{\delta \bar{E}}{\delta \phi_{\alpha}}\bigg)^\perp \; \; \forall \alpha \in [0, 1],
\end{equation}
where $M$ is a mobility coefficient, $\delta \bar{E} / \delta \phi_{\alpha}$ is the functional derivative of (\ref{E_modified}) evaluated on the image $\phi_{\alpha}$ and $(\delta \bar{E} / \delta \phi_{\alpha})^\perp$ is its component normal to the string. This last quantity can be computed as $(\delta \bar{E} / \delta \phi_{\alpha})^\perp = \delta \bar{E} / \delta \phi_{\alpha} - \Braket{\delta \bar{E} / \delta \phi_{\alpha} | \tau} \tau$, where $\tau = \partial_{\alpha} \phi_{\alpha} / \Braket{\partial_\alpha \phi_{\alpha} | \partial_\alpha \phi_{\alpha}}^{1/2}$ is the unit tangent to the string and $\Braket{\cdot|\cdot}$ is the standard $L_2$ inner product. In this way, at steady state, the string converges to a minimal energy path \cite{Cameron2011TheSM}. In order to eliminate the trouble of projecting the functional derivative and in order to use the equal arc-length parameterization, the string dynamics can be rewritten \cite{E2007SimplifiedAI} as
\begin{equation}
\dfrac{\partial \phi_{\alpha}}{\partial t} = - M \dfrac{\delta \bar{E}}{\delta \phi_{\alpha}} + \bar{\lambda} \tau \; \; \forall \alpha \in [0, 1],
\end{equation}
where $\bar{\lambda} = \lambda + M \Braket{\delta \bar{E} / \delta \phi_{\alpha} | \tau}$ and $\lambda$ is a Lagrange multiplier for the purpose of enforcing the chosen parameterization $\partial_\alpha \Braket{\partial_\alpha \phi_{\alpha} | \partial_\alpha \phi_{\alpha}}^{1/2} = 0$.

The algorithm follows the steps:
\begin{enumerate}
	\item Evolution from $t$ to $t + \Delta t$ of the discrete string, made up of $N$ images $\phi_i$, with the dynamics $$\dfrac{\partial \phi_i}{\partial t} = - M \dfrac{\delta \bar{E}}{\delta \phi_i} , \; \; i = 1, ... \, , N \, .$$ Time integration is performed in wave number space by means of the semi-implicit Euler single step scheme. The evolved images at time $t + \Delta t$ are denoted as $\tilde{\phi}_{i}$. 
	\item Computation of the arc lengths corresponding to the evolved images: 
	\begin{align*}
	 & s_0  =  0, \\ 
	 & s_i = s_{i-1} + \Braket{\tilde{\phi}_{i} - \tilde{\phi}_{i-1} | \tilde{\phi}_{i} - \tilde{\phi}_{i-1}}^{1/2} \, , \\
	 & i  =  1, ... \, , N \, . 
	\end{align*}
	Thus, the evolved images have parameters $\alpha_i = s_i / s_N$.
	\item Linear interpolation of the evolved images in order to compute the new images at equal arcs $\alpha_i = i/N$. These are the actual solutions at time $t+ \Delta t$. It is worth noticing that linear interpolation conserves vesicles volume.
	\item Go back to one and iterate until convergence.
\end{enumerate}

\subsection{Force fields computation}\label{Force_Field_Computation}

Given a membrane state, it is possible to compute the external force needed to balance the elastic force arising from the energy of the membrane. For this purpose, let's consider an arbitrary and infinitesimal variation $\delta \phi$ of the phase-field, consistent with the area and volume constraints, if present. This variation results in a spatial displacement $\delta \bm{x}$ of the field lines. The displacement can be thought to occur in a virtual time interval $\delta t$, within which the field lines move with a virtual velocity $\bm{u}$ such that $\partial \phi / \partial t = - \bm{\nabla} \phi \cdot \bm{u}$ (null material derivative condition). By integrating in time this last equation from $t$ to $t + \delta t$, we are left with the first order approximation
\begin{equation}
\delta \phi = - \bm{\nabla} \phi \cdot \bm{u} \delta t = - \bm{\nabla}\phi \cdot \delta \bm{x}.
\end{equation}
Hence, the work performed by the external force field $\bm{f}$ to deform the membrane is
\begin{equation}
	\begin{split}
		& \int_{\Omega} \bm{f} \cdot \delta \bm{x} \, dV  = \delta \bar{E} = \\
	    & \int_{\Omega} \dfrac{\delta \bar{E}}{\delta \phi} \, \delta \phi \, dV  = - \int_{\Omega} \dfrac{\delta \bar{E}}{\delta \phi} \bm{\nabla} \phi \cdot \delta \bm{x} \, dV,
	\end{split}
\end{equation}
and one can identify the force field
\begin{equation}
\bm{f} = - \dfrac{\delta \bar{E}}{\delta \phi} \bm{\nabla} \phi \; ,
\end{equation}
thanks to the arbitrariness of $\delta \bm{x}$.

\begin{acknowledgments}
	
Support is acknowledged from the 2020 Sapienza Large Project: Dynamics of Biological and Artificial Lipid Bilayer Membranes. Concerning computational resources we acknowledge: PRACE for awarding us access to Marconi successor at CINECA, Italy, PRACE 23rd call project Nr. 2021240074; DECI 17 SOLID project for resource Navigator based in Portugal at \url{https://www.uc.pt/lca/} from the PRACE aisbl; CINECA award under the ISCRA initiative, for the availability of high performance computing resources and support (ISCRA-B FHDAS).

\end{acknowledgments}

\appendix

\renewcommand{\appendixname}{APPENDIX}

\section{PHASE-FIELD VERSION OF THE GAUSS-BONNET THEOREM}\label{Gauss_Bonnet_appendix}

Let's assume that
\begin{equation}
	\phi(\bm{x}) = \tanh\bigg(\dfrac{d(\bm{x})}{\epsilon \sqrt{2}}\bigg),
\end{equation}
where $\bm{x} \in \Omega$, being $\Omega$ a cylindrical domain of radius $R$ and height $L$ in the ordinary three-dimensional space, and $d(\cdot)$ the signed distance from an axisymmetric surface in $\Omega$. This assumption leads to
\begin{equation}\label{Gradphi_phi2}
	|\bm{\nabla} \phi| = \dfrac{(1-\phi^2)}{\epsilon \sqrt{2}} \; ,
\end{equation}
and, moreover, we set
\begin{equation}
h(\phi) = \bigg(\dfrac{1-\phi^2}{\epsilon \sqrt{2}}\bigg)^4 \; \; .
\end{equation}
Using the cylindrical coordinates system, it is possible to show by a direct computation \cite{Du2005RetrievingTI} that one of the two principal curvatures is
\begin{equation}
k_1 = -\dfrac{\partial_r \phi}{r |\bm{\nabla} \phi|} \; \; .
\end{equation}
Therefore, remembering that $\bm{\nabla} \cdot \bm{n} = -(k_1 + k_2)$ and $\bm{n} \cdot \bm{\nabla} k_i = k_i^2$, with $\bm{n} = \bm{\nabla}d$, Eq.~(\ref{EGf}) can be rewritten as
\begin{align*}
&E_G[\phi] \; = \\
& = \; k_G \dfrac{35}{8 \sqrt{2}} \epsilon^3 \int_{\Omega} h(\phi) \, k_1 k_2 \, dV \; = \\
& = \; - k_G \dfrac{35}{8 \sqrt{2}} \epsilon^3 \int_{\Omega} h(\phi) \, \bm{\nabla} \cdot (\bm{n}  \, k_1) \, dV \; = \\
& = \; k_G \dfrac{35}{8 \sqrt{2}} \epsilon^3 \int_{\Omega} \dfrac{d h}{d \phi} \, \bm{\nabla} \phi \cdot \bm{n}  \, k_1 \, dV \, + \, I_{\partial \Omega} \; = \\
& = \; k_G \dfrac{35}{8 \sqrt{2}} \epsilon^3 \int_{\Omega} \dfrac{d h}{d \phi} \, |\bm{\nabla} \phi| \, k_1 \, dV \, + \, I_{\partial \Omega} \; = \\
& = \; - k_G \dfrac{35}{4 \sqrt{2}} \epsilon^3 \pi \int_{-L/2}^{+L/2} dz \, \int_{0}^{R} \dfrac{d h}{d \phi} \, \dfrac{\partial \phi}{\partial r} \, dr \, + \, I_{\partial \Omega}\; = \\
& = k_G \dfrac{35}{4 \sqrt{2}} \epsilon^3 \pi \int_{-L/2}^{+L/2} [h(\phi(r=0, z)) -  h(\phi(r = R, z))] dz \, + \\
& \; \; \; \, + I_{\partial \Omega} \; ,
\end{align*}
where 
$$I_{\partial \Omega} = - \, k_G \dfrac{35}{8 \sqrt{2}} \epsilon^3 \int_{\partial \Omega} h(\phi) \, k_1 \, \bm{n}_{\Omega} \cdot \bm{n} \, dS \; .$$
Supposing to have a single, connected, closed surface, after letting $\Omega$ invade $\mathbb{R}^3$, and still considering relation (\ref{delta2}) with (\ref{Gradphi_phi2}), we obtain
\begin{align*}\label{Gauss_Bonnet_PF}
&\lim_{\epsilon \to 0} E_G[\phi] \; = \\
& = \; 2 \pi k_G \int_{-\infty}^{+\infty} \delta(d(r=0, z)) \, dz \; = \\
& = \; 4 \pi k_G \, (1-g) \;,
\end{align*}
recovering the Gauss-Bonnet theorem (\ref{Gauss_Bonnet_theorem}) in the axially-symmetric case. The last equality is 
justified by the fact that the Dirac delta function counts the intersections of the surface with the z-axis, which is equivalent to checking whether the surface has a hole.

\section{PHASE-FIELD MODEL NUMERICAL VALIDATION}\label{Numerical_validation}

\begin{figure}[h]
	\includegraphics[width=8.6cm]{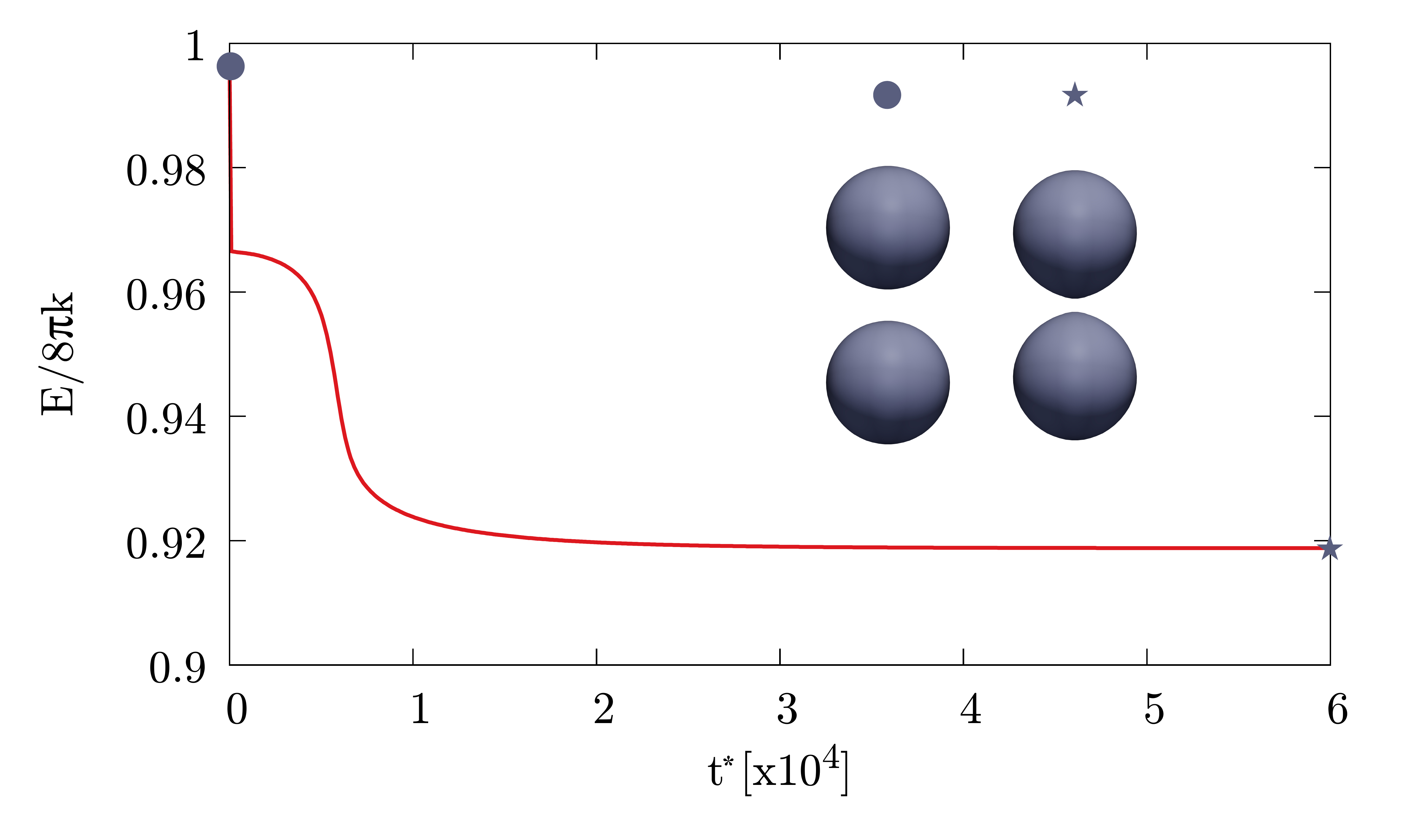}
	\caption{{\small Two spheres of radius $R^* = 10$ and $R^*/2$ distant from each other do not merge during the Allen-Cahn dynamics in presence of the new phase-field Gaussian energy term. For this simulation we used a $[0, 40] \times [0, 40] \times [0, 66]$ full 3D $x^*-y^*-z^*$ domain with a grid of $40 \times 40 \times 66$ nodes, $\epsilon^* = h^* = 1$, $m^* = 0$, $1/\lambda = 20 \sqrt{2}$, $M^* = 1$ and time step $dt^* = 0.8$.}}
	\label{Sfere_validazione}
\end{figure}

In literature \cite{Esedoglu2012CollidingII}, it is well known that, in the presence of the bending energy alone, two initially close-by spheres merge together during the Allen-Cahn dynamics, Eq. (\ref{Allen-Cahn}). If the area and volume constraints are included, at the steady state, a dumbbell shape with a reduced volume $v = 1/\sqrt{2}$ is obtained. This happens because the bending energy of the two spheres is greater than that of the obtained dumbbell shape. Moreover, such a numerical experiment shows that there is no energy barrier for the process. This behavior is no longer possible if Gaussian energy is also included. Indeed, in this case, the whole Canham-Helfrich energy of two spheres is less than that of the dumbbell shape. Therefore, our first numerical validation experiment is to repeat this simulation including the new Gaussian energy term, Eq.~(\ref{Gauss}). As shown in Fig.~\ref{Sfere_validazione}, two spheres of equal radius $R^* = 10$ at distance $R^*/2$ from each other do not merge. This simulation has been carried out in a $[0, 40] \times [0, 40] \times [0, 66]$ full 3D $x^*-y^*-z^*$ domain with a grid of $40 \times 40 \times 66$ nodes (grid length interval $h^* = 1$), $\epsilon^* = h^* = 1$, $1/\lambda = 20 \sqrt{2}$, $M^* = 1$ and time step $dt^* = 0.8$. In Fig.~\ref{Sfere_validazione}, the energy monotonically decreases over time, revealing the stability of the scheme. At steady state, the final computed bending and Gaussian energies are $E_B^* \approx 1.941$ and $E_G^* \approx -1.023$. The same simulation has also been carried out  in the $r^*-z^*$ plane, exploiting the axisymmetry,
with $\epsilon^* = 2h^* = 1$ and two different time steps, $dt^* = 0.8$ and $dt^* = 0.4$, respectively, still obtaining the same behavior. Convergence has also been observed setting $\epsilon^* = 1.5 \, h^* = 1$ and $1/\lambda = 40 \sqrt{2}$, with the two spheres at distance $R^*/4$ from each other.

\begin{figure}[h]
	\includegraphics[width=8.6cm]{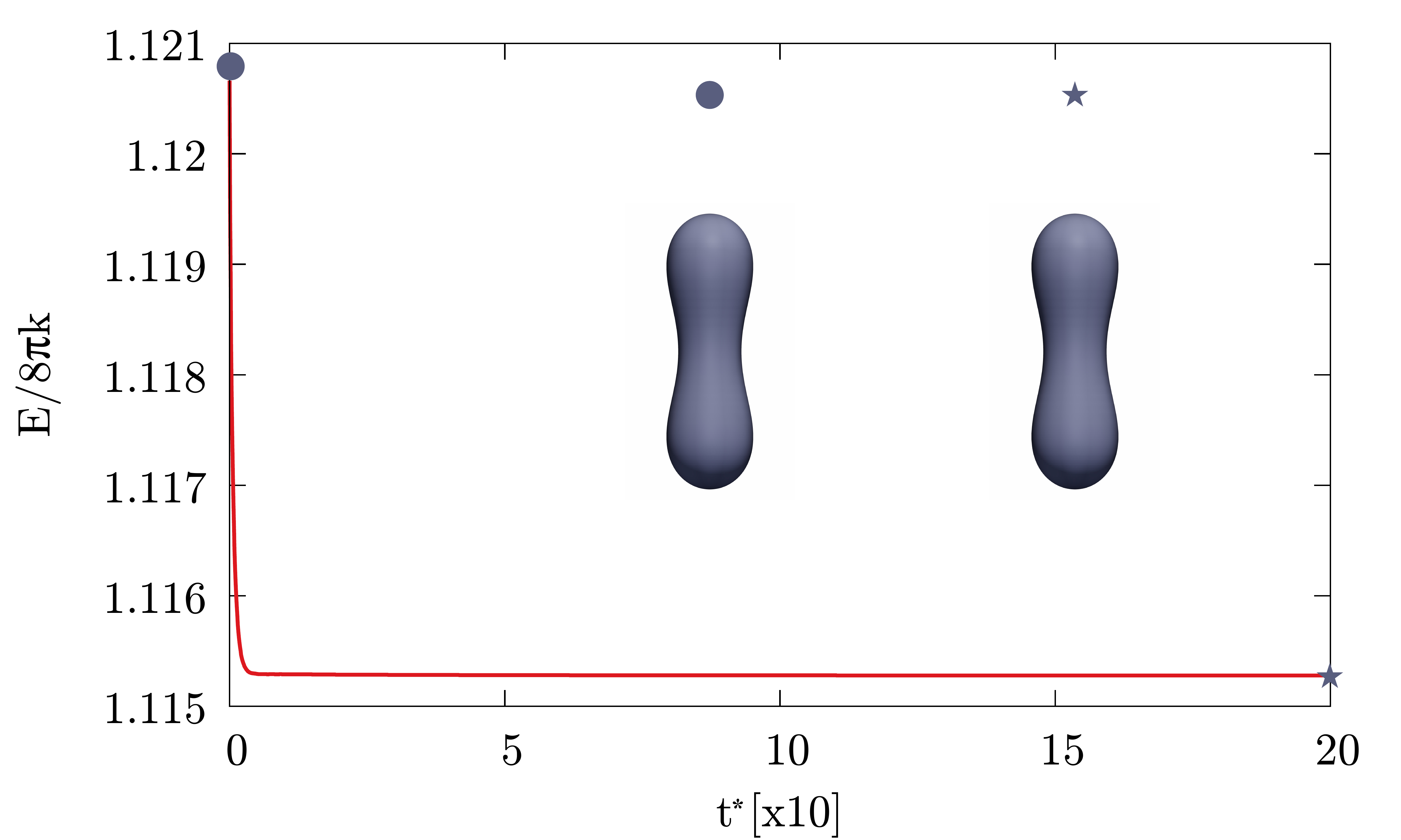}
	\caption{{\small A dumbbell shape, which is a minimal configuration for the bending energy alone, is also a minimal configuration for the whole Canham-Helfrich energy. This z-axial symmetric simulation has been carried out in a $[0, 28] \times [0, 120]$ computational domain in the $r^*-z^*$ plane with a $42 \times 180$ mesh, $\epsilon^* = 1.5 h^*$, $m^* = 0$, $1/\lambda = 40 \sqrt{2}$, $M^* = 8$ and $dt = 0.01$.}}
	\label{Prolato_validazione}
\end{figure}

With the same parameters, let's take the dumbbell shape obtained merging the two spheres in a simulation with the sole bending energy, and let's use it as a new initial condition for the Allen-Cahn dynamics where the Gaussian energy is now included. As shown in Fig.~\ref{Prolato_validazione}, the dumbbell shape remains substantially unchanged, showing that the configuration is still a local energy minimum and that there exists an energy barrier that prevents it from dividing into two spheres. It is worth noticing that the computed energies are in excellent agreement with the ones reported in \cite{Seifert1991ShapeTO}. Bending and Gaussian contributions to the energy of the final configuration are $E_B^* \approx 1.625$ and $E_G^* \approx -5.095 \cdot 10^{-1}$.

\begin{figure}[H]
	\includegraphics[width=8.6cm]{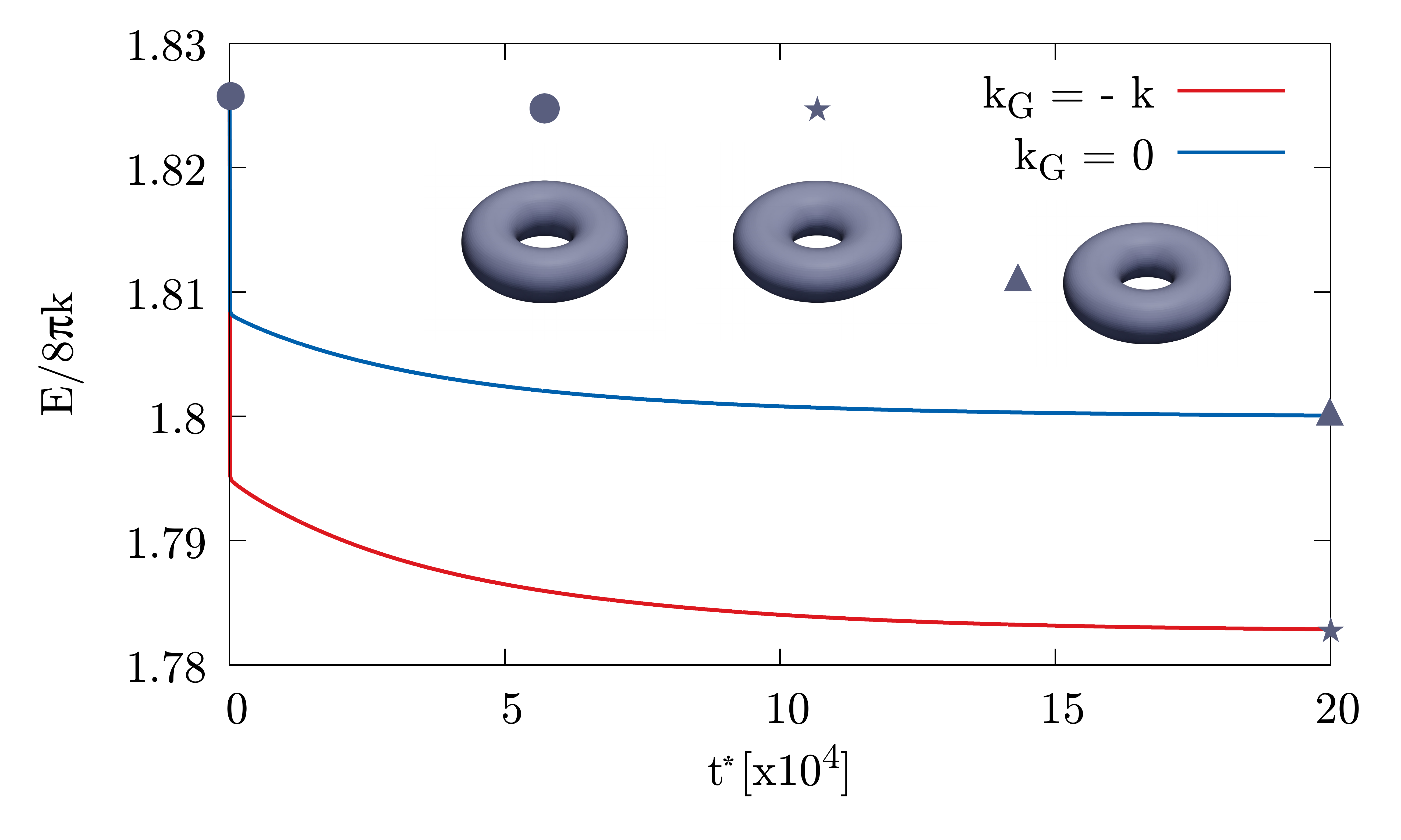}
	\caption{{\small A torus with exact circular cross section of radius $R^*=10$ evolves to a torus with a cross section that is not perfectly circular. This happens both with $k_G = 0$ (top blue line) and $k_G = - k$ (bottom red line). These z-axial symmetric simulations have been carried out in a $[0, 40] \times [0, 40]$ computational domain in the $r^*-z^*$ plane with a $60 \times 60$ mesh, $\epsilon^* = 1.5 h^* = 1$, $m^* = 0$, $1/\lambda = 20 \sqrt{2 \pi}$, $M^* = 1$ and $dt^* = 1$.}}
	\label{Toro_validazione}
\end{figure}

Finally, we test a toroidal topology case. The initial condition is a torus with exact circular cross section of radius $R^*=10$ and $v \approx 0.6$. The dynamics leads to a torus with a cross section that is no more perfectly circular, in excellent agreement with \cite{Seifert1991VesiclesOT}, both as regards the shape and the energy. Figure~\ref{Toro_validazione} shows the energy evolution both with and without the Gaussian energy term. The two dynamics appear to be very similar, confirming that the Gaussian energy term plays no role as long as no topological transitions occur. These axisymmetric simulations have been carried out in a $[0, 40] \times [0, 40]$ computational domain in the $r^*-z^*$ plane with a grid of $60 \times 60$ nodes, $\epsilon^* = 1.5 \, h^* = 1$, $1/\lambda = 20 \sqrt{2 \pi}$, $M^* = 1$ and $dt^* = 1$. With the Gaussian term included, the final computed bending and Gaussian energies  are $E_B^* \approx 1.831$ and $E_G^* \approx -4.813 \cdot 10^{-2}$, respectively. Noteworthy, the computed Gaussian energy is greater than that reported in Table~\ref{Table_gauss}, since higher order corrections to the $\tanh\text{-solution}$ are present, see Section~\ref{PFMD}, Eq~(\ref{EGf}). Performing the same simulation with $1/ \lambda = 40 \sqrt{2 \pi}$, $\epsilon^* = 1.5 \, h^* = 1$, and the same $dt^*$, the computed energies at the same final time are $E_B^* \approx 1.813$ and $E_G^* \approx -1.096 \cdot 10^{-2}$. The error with respect to the data reported in Table~\ref{Table_gauss} decreases, since, by reducing the dimensionless thickness $\lambda$, the higher order terms become less and less important.

In all simulations presented in this Appendix, vesicles area and volume are conserved with the same accuracy reported in Fig.~\ref{fig:stringpanel}.

\section{FUNCTIONAL DERIVATIVES}\label{Appendix_functional_derivatives}

The functional derivative of the energy (\ref{Energy}) is
\begin{equation}
	\dfrac{\delta E}{\delta \phi} = \dfrac{\delta E_B}{\delta \phi} + \dfrac{\delta E_G}{\delta \phi}.
\end{equation}
The bending term is well known in literature and its explicit expression is
\begin{equation}
	\dfrac{\delta E_B}{\delta \phi} = k \, \dfrac{3}{2 \sqrt{2}} \, \epsilon \, \bigg[\nabla^2 \psi_B - \dfrac{\psi_B}{\epsilon^2} (3\phi^2 - 1 + 2 \sqrt{2} \epsilon m \phi ) \bigg].
\end{equation}
As regards the Gaussian term, noticing that
\begin{equation}
	\bm{\nabla} \cdot \bigg(\bm{\nabla} \phi \nabla^2 \phi - \dfrac{\bm{\nabla}|\bm{\nabla} \phi|^2}{2} \bigg) = (\nabla^2 \phi)^2 + \bm{\nabla} \phi \cdot \bm{\nabla} \nabla^2 \phi - \dfrac{\nabla^2 |\bm{\nabla} \phi|^2}{2},
\end{equation}
an integration by parts of (\ref{Gauss}) leads to
\begin{equation}
	E_{G}[\phi] = k_{G} \, \dfrac{35}{8 \sqrt{2}} \, \epsilon^3 \, \int_{\Omega} \tilde{\psi}_G  \; dV \; ,
\end{equation}
\begin{equation}
	\tilde{\psi}_G = \dfrac{\bm{\nabla} |\bm{\nabla} \phi|^2 \cdot \bm{\nabla} |\bm{\nabla} \phi|^2}{2} - (\bm{\nabla} |\bm{\nabla} \phi|^2 \cdot \bm{\nabla} \phi) \nabla^2 \phi \, .
\end{equation}\\
This simplifies the computation of the functional derivative, which turns out to be
\begin{equation}
\begin{split}
\dfrac{\delta E_G}{\delta \phi} = & \; k_{G} \, \dfrac{35}{8 \sqrt{2}} \, \epsilon^3 \, \bigg\{ 2 \bm{\nabla} \cdot [(\nabla^2 |\bm{\nabla} \phi|^2) \bm{\nabla} \phi] \\
& + \bm{\nabla} \cdot(\nabla^2 \phi \bm{\nabla} |\bm{\nabla} \phi|^2) - \nabla^2(\bm{\nabla}|\bm{\nabla} \phi|^2 \cdot \bm{\nabla} \phi) \\
& - 2 \bm{\nabla} \cdot [\bm{\nabla} \phi \bm{\nabla} \cdot(\nabla^2 \phi \bm{\nabla} \phi)] \bigg\}. \\
\end{split}
\end{equation}
In a more readable form:
\begin{equation}
\begin{split}
\dfrac{\delta E_G}{\delta \phi}  =  & \; k_{G} \, \dfrac{105}{2 \sqrt{2}} \, \epsilon^3 \, \bigg[ \dfrac{\partial^2 \phi}{\partial x^2}\bigg(\dfrac{\partial^2 \phi}{\partial y \partial z}\bigg)^2 \\
& + \; \dfrac{\partial^2 \phi}{\partial y^2}\bigg(\dfrac{\partial^2 \phi}{\partial x \partial z}\bigg)^2 + \, \dfrac{\partial^2 \phi}{\partial z^2}\bigg(\dfrac{\partial^2 \phi}{\partial x \partial y}\bigg)^2 \\ 
& - \; \dfrac{\partial^2 \phi}{\partial x^2}\dfrac{\partial^2 \phi}{\partial y^2}\dfrac{\partial^2 \phi}{\partial z^2} \, - \, 2\dfrac{\partial^2 \phi}{\partial x \partial y}\dfrac{\partial^2 \phi}{\partial x \partial z}\dfrac{\partial^2 \phi}{\partial y \partial z} \bigg], \\
\end{split}
\end{equation}
which further simplifies  in the axisymmetric case to
\begin{equation}
\begin{split}
\dfrac{\delta E_G}{\delta \phi}  =  k_{G} \, \dfrac{105}{2 \sqrt{2}} \, \epsilon^3 \, \dfrac{1}{r}\dfrac{\partial \phi}{\partial r} \, \bigg[\bigg(\dfrac{\partial^2 \phi}{\partial r \partial z}\bigg)^2  - \dfrac{\partial^2 \phi}{\partial r^2}\dfrac{\partial^2 \phi}{\partial z^2} \bigg].
\end{split}
\end{equation}
Furthermore, the functional derivatives of the area (\ref{area}) and volume (\ref{volume}) are
\begin{equation}
	\dfrac{\delta A}{\delta \phi} = \dfrac{3}{2 \sqrt{2}} \, \epsilon \, \bigg[\dfrac{1}{\epsilon^2} \phi (\phi^2 -1) - \nabla^2 \phi \bigg],
\end{equation}
\begin{equation}
	\dfrac{\delta V}{\delta \phi} = \dfrac{1}{2} .
\end{equation}
In conclusion, the functional derivative of the augmented energy (\ref{E_modified}) is
\begin{equation}
\begin{split}
& \dfrac{\delta \bar{E}}{\delta \phi} = \dfrac{\delta E_B}{\delta \phi} + \dfrac{\delta E_G}{\delta \phi} + \gamma \dfrac{\delta A}{\delta \phi} + \Delta p \dfrac{\delta V}{\delta \phi} + \\
& + M_1(A[\phi] - A_0) \dfrac{\delta A}{\delta \phi} + M_2(V[\phi] - V_0) \dfrac{\delta V}{\delta \phi} \; .
\end{split}
\end{equation}

~
~

\onecolumngrid
~\newline
~\newline
\twocolumngrid

%


\end{document}



\title{SUPPLEMENTAL MATERIAL \\ Topological transitions in fluid lipid vesicles: \\ activation energy and force fields}
\author{Matteo Bottacchiari, Mirko Gallo, Marco Bussoletti}%
\author{Carlo Massimo Casciola}
\email{carlomassimo.casciola@uniroma1.it}
\affiliation{Department of Mechanical and Aerospace Engineering, Sapienza Universit\`a di Roma, Rome, Italy}%

\date{\today}

\pacs{}

\maketitle

\section{Asymptotic analysis}\label{Asymptotic_Analysis}

The phase-field $\phi(\bm{x})$ is a regular function, defined on a domain $\Omega \in \mathbb{R}^3$, which takes values in the range $(-1,1)$.  An energy given by the functional $E[\phi]$ is attached to each field configuration, as  specified by the equations

\begin{equation}\label{Energy}
	E[\phi] = E_B[\phi] + E_G[\phi],
\end{equation}
with
\begin{equation}\label{Bending}
	E_{B}[\phi] = k \, \dfrac{3}{4 \sqrt{2}} \, \epsilon \, \int_{\Omega} \Psi_{B}^2 \; dV,
\end{equation}
\
\begin{equation}\label{PsiB}
	\Psi_{B} = \nabla^2 \phi - \dfrac{1}{\epsilon^2} (\phi^2 - 1) (\phi +  \sqrt{2}\epsilon m)
\end{equation}
and
\begin{equation}\label{Gauss}
	E_{G}[\phi] = k_{G} \, \dfrac{35}{16 \sqrt{2}} \, \epsilon^3 \, \int_{\Omega} \Psi_{G} \; dV,
\end{equation}
\
\begin{equation}\label{PsiG}
\Psi_{G} = \dfrac{ \bm{\nabla} |\bm{\nabla} \phi|^2 \cdot \bm{\nabla} |\bm{\nabla} \phi|^2}{2} - (\bm{\nabla} |\bm{\nabla} \phi|^2 \cdot \bm{\nabla} \phi) \nabla^2 \phi \; + \; |\bm{\nabla} \phi|^2 \bigg[(\nabla^2 \phi)^2 + \bm{\nabla} \phi \cdot \bm{\nabla} \nabla^2 \phi - \dfrac{\nabla^2 |\bm{\nabla} \phi|^2}{2}\bigg].
\end{equation}

In the limit of vanishing $\epsilon$ the field energy  will be shown below to converge to the sharp interface energy of the Canham-Helfrich model, see Eq.~(1) in the main text, for a membrane identified with the mid surface implicitly defined by the field isosurface  $\phi(\bm{x}) = 0$. In this case,  the various parameters entering the expression for the energy will be interpreted as follows: $m$ is the spontaneous curvature of the membrane, assumed to be closed, taken to be positive if the membrane bulges towards the exterior; $k$ is the bending rigidity; $k_G$ is the Gaussian curvature modulus (for definiteness,  $k_G = -k$ will be assumed, as suggested by the experimental data, see the main text for details). Hence the equilibrium configurations of a Canham-Helfrich membrane can be obtained either by minimizing the Canham-Helfrich functional with respect to the membrane configuration or, alternatively, by minimizing the phase-field energy functional in the limit of vanishing $\epsilon$. As explained in the main text, apart from practical advantages in the minimization process, the phase-field approach presents the main advantage over the sharp interface description of allowing for the membrane topological transitions.

More technically, it is shown below that an ansatz for the phase function of the form
\begin{equation}\label{ansatz}
	\phi(\bm{x}) = f\bigg(\dfrac{d(\bm{x})}{\epsilon}\bigg),
\end{equation} 
where $d(\cdot)$ is the signed distance from the (assigned) membrane midsurface configuration 
$\Gamma = \left\{\bm{x}: \phi(\bm{x}) = 0 \right\}$ and $f$ is such that $f(\pm \infty) = \pm 1$ and $f(0) = 0$:
i) does indeed allow to find a minimizer of the field energy for small enough $\epsilon$ and given midsurface geometry; ii) leads to a limiting, sharp interface form of the field energy  which coincides with the Canham-Helfrich energy, which is a functional of the membrane configuration. From the discussion, it will become clear that the parameter $\epsilon > 0$ is related to the width of the diffuse interface and, thus, to the small, but finite, membrane thickness. 

We choose to define the signed distance such that $\bm{n} = \bm{\nabla} d$ computed on $\Gamma$ is the inward-pointing unit normal to the vesicle. It is worth noticing that the signed distance function  satisfies the eikonal equation, namely $|\bm{n}| = 1$. As anticipated, setting $d^*(\bm{x}) = d(\bm{x}) / \epsilon$, we also require that $\lim_{d^* \to \pm \infty} \phi = \pm 1$ and $\phi = 0$ for $d=0$. Therefore, $\pm 1$ are the values for the stable phases of the inside and outside bulk and the level set $\phi = 0$ identifies the membrane midsurface. 

After introducing the membrane characteristic length $D_{ve} = \sqrt{A/\pi}$, where $A$ is a characteritic  membrane area, the sharp-interface limit corresonds to  $\lambda = \epsilon / D_{ve} \to 0$. 
Since  $\bm{\nabla} \phi (\bm{x}) = f'(d^*(\bm{x})) \, \bm{n}/\epsilon$, where the prime denotes the derivative with respect to $d^*(\bm{x})$, a direct computation leads to
\begin{equation}
	E_{B}[\phi] = \; k \, \dfrac{3}{4 \sqrt{2}} \, \lambda \, \int_{\bar{\Omega}} \bigg[ \dfrac{1}{\lambda^2}\bigg(f'' - (f^2-1)f\bigg) \; + \; \dfrac{1}{\lambda}\bigg(f' \, \bm{\bar{\nabla}} \cdot \bm{n} + (1-f^2)\sqrt{2}\bar{m}\bigg) \bigg]^2 \, d\bar{V},
\end{equation}
\
\begin{equation}
	E_{G}[\phi] = k_G \, \dfrac{35}{16 \sqrt{2}} \, \int_{\bar{\Omega}} \dfrac{f'^4}{\lambda} \bigg[ (\bm{\bar{\nabla}} \cdot \bm{n})^2 + \bm{n} \cdot \bm{\bar{\nabla}} (\bm{\bar{\nabla}} \cdot \bm{n}) \bigg]  \, d\bar{V},
\end{equation}
where the bar indicates lengths normalized with $D_{ve}$.
Following \cite{Wang2008AsymptoticAO}, in order to minimize $E = E_{B} + E_{G}$ as $\lambda \to 0$ the phase-field function is expanded as 
\begin{equation}
\label{expansion}
\phi(\bm{x}) = f(d^*(\bm{x})) = f_{0}(d^*(\bm{x})) +  \lambda f_{1}(d^*(\bm{x})) + R_{\lambda}(d^*(\bm{x}))
\end{equation}
where $R_{\lambda}(d^*(\bm{x})) = O(\lambda^2)$, obtaining
\begin{equation}
	\begin{split}
	& E_B[\phi] = \; k \, \dfrac{3}{4 \sqrt{2}} \, \lambda \, \int_{\bar{\Omega}} \bigg[ \dfrac{1}{\lambda^2}\bigg(f_0'' - (f_{0}^2-1)f_{0}\bigg) \; + \\ & + \; \dfrac{1}{\lambda} \bigg( f_1'' - f_1(3f_{0}^2 - 1) + f_{0}' \, \bm{\bar{\nabla}} \cdot \bm{n} + (1-f_{0}^2)\sqrt{2}\bar{m}\bigg) + h \bigg]^2 \, d\bar{V} \; ,
	\end{split}
\end{equation}
\
\begin{equation}
	E_{G}[\phi] = k_G \, \dfrac{35}{16 \sqrt{2}} \, \int_{\bar{\Omega}} \bigg(\dfrac{f_{0}'^4}{\lambda} + 4 f_{0}'^3 f_1' + \lambda g \bigg) \bigg[ (\bm{\bar{\nabla}} \cdot \bm{n})^2 + \bm{n} \cdot \bm{\bar{\nabla}} (\bm{\bar{\nabla}} \cdot \bm{n}) \bigg]  \, d\bar{V} \; ,
\end{equation}
with $g = g(d^*(\bm{\bar{x}}))$ and $h$ a combination of functions factorized in the form $\eta(d^*(\bm{\bar{x}})) \xi(\bm{\bar{x}})$. Therefore, both $g$ and $h$ are $O(1)$ in $\lambda$.
Expansion~(\ref{expansion}) entails the ordering of the two partial energy functionals as power series in the small 
parameter $\lambda$.
Therefore, in order to minimize the otherwise dominating bending energy (order $1/\lambda^3$), the leading order term $f_0$ must satisfy
\begin{equation}\label{eq_tanh}
	f_{0}'' = (f_{0}^2 - 1)f_{0},
\end{equation}
i.e., 
\begin{equation}\label{f0}
	f_{0}(d^*(\bm{x})) = \tanh\bigg(\dfrac{d(\bm{x})}{\epsilon \sqrt{2}}\bigg)\, ,
\end{equation}
which identifies  $\epsilon$ as directly proportional to the interface width.  
Note that, once $f_0$ is determined, the order $\lambda^{-1}$ contribution to the Gaussian energy is fixed. Concerning the bending energy, we are left with
\begin{equation*}
E_{B}[\phi] = \; k \, \dfrac{3}{4 \sqrt{2}} \, \lambda \, \int_{\bar{\Omega}} \bigg[\dfrac{1}{\lambda} \bigg( f_1'' - f_1(3f_{0}^2 - 1) + f_{0}' \, \bm{\bar{\nabla}} \cdot \bm{n} + (1-f_{0}^2)\sqrt{2}\bar{m}\bigg) + h \bigg]^2 \, d\bar{V} \; =
\end{equation*}
\
\begin{equation}\label{EB1}
k \dfrac{3}{4 \sqrt{2}} \dfrac{1}{\lambda} \int_{\bar{\Omega}} \bigg[ \big[f_1'' - f_1(3f_{0}^2 - 1) \big]^2 + 2 f_{0}' \big[f_1'' - f_1(3f_{0}^2 - 1) \big] \big[\bm{\bar{\nabla}} \cdot \bm{n} + 2\bar{m}\big] + {f_{0}'}^2 \big[\bm{\bar{\nabla}} \cdot \bm{n} + 2\bar{m}\big]^2 + \lambda \tilde{h} \bigg] dV,
\end{equation}
where we have used 
\begin{equation}\label{Gradient_tanh}
	\sqrt{2} f_{0}' = (1 - f_{0}^2) \ ,
\end{equation}
and $\tilde{h} = O(1)$ shares the same structure of $h$.

Now, considering that
\begin{equation}\label{Third_derivative_tanh}
	f_0''' = f_0'(3{f_0}^2 - 1) \ ,
\end{equation}
one finds that
\begin{equation}
	\int_{\bar{\Omega}} 2 f_{0}' \big[f_1'' - f_1(3f_{0}^2 - 1) \big] \big[\bm{\bar{\nabla}} \cdot \bm{n} + 2\bar{m}\big] \, d\bar{V} \; = \; \int_{\bar{\Omega}} 2 \big[f_0' f_1'' - f_0''' f_1\big] \big[\bm{\bar{\nabla}} \cdot \bm{n} + 2\bar{m}\big] \, d\bar{V}.
\end{equation}
An integration by parts leads to
\begin{align*}
    & 2 \int_{\bar{\Omega}} \big[f_1 f_0''' \big] \big[\bm{\bar{\nabla}} \cdot \bm{n} + 2\bar{m}\big] \, d\bar{V} = \; 2 \lambda \int_{\bar{\Omega}} \big[f_1 \bm{\bar{\nabla}} f_0'' \cdot \bm{n} \big] \big[\bm{\bar{\nabla}} \cdot \bm{n} + 2\bar{m}\big] \, d\bar{V} = \\
    & = \; - 2 \int_{\bar{\Omega}} \big[f_1' f_0'' \big] \big[\bm{\bar{\nabla}} \cdot \bm{n} + 2\bar{m}\big] \, d\bar{V} \; - \; 2 \lambda \int_{\bar{\Omega}} \big[f_1 f_0'' \big] \bm{\bar{\nabla}} \cdot \big[ \bm{n} \big(\bm{\bar{\nabla}} \cdot \bm{n} + 2\bar{m} \big)  \big] \, d\bar{V} = \\
    & = \; 2 \int_{\bar{\Omega}} \big[f_1'' f_0' \big] \big[\bm{\bar{\nabla}} \cdot \bm{n} + 2\bar{m}\big] \, d\bar{V} \; + \; 2 \lambda \int_{\bar{\Omega}} \big[f_1' f_0' - f_1 f_0'' \big] \bm{\bar{\nabla}} \cdot \big[ \bm{n} \big(\bm{\bar{\nabla}} \cdot \bm{n} + 2\bar{m} \big)  \big] \, d\bar{V}
\end{align*}
and therefore
\begin{equation*}
	\int_{\bar{\Omega}} 2 f_{0}' \big[f_1'' - f_1(3f_{0}^2 - 1) \big] \big[\bm{\bar{\nabla}} \cdot \bm{n} + 2\bar{m}\big] \, d\bar{V} \; = - \; \int_{\bar{\Omega}} 2 \lambda \big[f_1' f_0' - f_1 f_0'' \big] \bm{\bar{\nabla}} \cdot \big[ \bm{n} \big(\bm{\bar{\nabla}} \cdot \bm{n} + 2\bar{m} \big)  \big] \, d\bar{V}.
\end{equation*}
Since the contribution of the above integral is of higher order,  substitution in (\ref{EB1}) shows that minimizing the $\lambda^{-1}$-order bending energy term is tantamount to solving the equation
\begin{equation}\label{f1_equation}
	f_1'' - (3f_{0}^2 - 1)f_1 = 0 .
\end{equation}

Two independent solutions are easily found. Given Eq.~(\ref{Third_derivative_tanh}), $f_{1a} = f_0'$ is clearly one of them, while the other can be sought in the form $f_{1b}(d^*) = w(d^*) f_0'(d^*)$. Thus, Equation~(\ref{f1_equation}) implies
\begin{equation}
	f_{0}' w'' + 2 f_{0}'' w' = 0 \, ,
\end{equation}
which, after setting $w' = z$ and still considering Eqs.~(\ref{eq_tanh})~ and~(\ref{Gradient_tanh}), is reduced to
\begin{equation}
	z' = 2 \sqrt{2} f_{0} \, z \ .
\end{equation}
The solution is $z(d^*) = \cosh^4 \bigg( \dfrac{d^*}{\sqrt{2}} \bigg)$, i.e.
\begin{equation}
	w(d^*) = \dfrac{1}{32} \big[ 12d^* + 8\sqrt{2} \sinh(\sqrt{2} \, d^*)) + \sqrt{2} \sinh(2\sqrt{2} \, d^*) \big] \ ,
\end{equation}
and therefore
\begin{equation}
	f_{1b}(d^*) = \dfrac{1}{32 \sqrt{2}} \bigg[ 12d^* + 8\sqrt{2} \sinh(\sqrt{2} \, d^*)) + \sqrt{2} \sinh(2\sqrt{2} \, d^*) \bigg] \sech^2 \bigg( \dfrac{d^*}{\sqrt{2}} \bigg) 
\end{equation}
($f_0'(d^*) = \sech^2(d^*/\sqrt{2})/\sqrt{2}$), where, due to the factors $\sqrt{2}$ in the arguments, $\lim_{d^* \rightarrow \pm \infty } f_{1b}(d^*) = \pm \infty$.
The two constants in the general solution of (\ref{f1_equation}),
\begin{equation}
	f_{1} = A f_{1a} + B f_{1b},
\end{equation}
are determined as follows. Since $f_{0}(\pm \infty) = \pm 1$, $B = 0$ by requiring that $\lim_{d^* \to \pm\infty} \phi = \pm 1$, and, given that $f_0(0) = 0$, $A = 0$ by requiring that $\phi = 0$ on the membrane midsurface. Therefore, $f_1 \equiv 0$, and we are left with
\begin{equation}\label{Asymptotic_EB}
	E_{B}[\phi] = k \dfrac{3}{4 \sqrt{2}} \int_{\bar{\Omega}} \dfrac{f_{0}'^2}{\lambda} (\bm{\bar{\nabla}} \cdot \bm{n} + 2\bar{m})^2 \, d\bar{V} \, + \, O(\lambda),
\end{equation}
\
\begin{equation}\label{Asymptotic_EG}
	E_{G}[\phi] = k_G \, \dfrac{35}{16 \sqrt{2}} \, \int_{\bar{\Omega}} \dfrac{f_{0}'^4}{\lambda} \bigg[ (\bm{\bar{\nabla}} \cdot \bm{n})^2 + \bm{n} \cdot \bm{\bar{\nabla}} (\bm{\bar{\nabla}} \cdot \bm{n}) \bigg]  \, d\bar{V} \, + \, O(\lambda^2),
\end{equation}
where the terms $O(\lambda)$ and $O(\lambda^2)$ in Eqs. (\ref{Asymptotic_EB}) and (\ref{Asymptotic_EG}), respectively, follow from the property $\int_{\bar{\Omega}} \eta(d^*(\bar{x})) \xi(\bar{x}) \, d\bar{V} = O(\lambda)$, for each $\eta(d^*(\bar{x})) = O(1)$ such that $\eta(\pm \infty) = 0$ sufficiently fast.
Denoting with $k_1$ and $k_2$ the principal curvatures of the membrane, we have $\bm{\nabla} \cdot \bm{n} = -(k_1 + k_2) = -2M$ and $\bm{n} \cdot \bm{\nabla} k_i = k_i^2$, with the result that $(\bm{\nabla} \cdot \bm{n})^2 + \bm{n} \cdot \bm{\nabla} (\bm{\nabla} \cdot \bm{n}) = 2 k_1 k_2 = 2G$. Now, noticing that for $\lambda \to 0$
\begin{equation}\label{delta}
	\dfrac{f_{0}'^2(\bar{d}(\bm{x})/\lambda)}{\lambda} \stackrel{\mathcal{W}}{\longrightarrow} \; \dfrac{2 \sqrt{2}}{3} \, \delta(\bar{d}(\bm{x})),
\end{equation}
\begin{equation}\label{delta2}
	\dfrac{f_{0}'^4(\bar{d}(\bm{x})/\lambda)}{\lambda} \stackrel{\mathcal{W}}{\longrightarrow} \; \dfrac{8 \sqrt{2}}{35} \, \delta(\bar{d}(\bm{x})),
\end{equation}
where the limits should be understood in a weak sense and $\delta(x)$ is the Dirac delta function, 
we are left with
\begin{equation}
	\lim_{\lambda \to 0} E[\phi] \; = \; 2k \int_{\Gamma} (M - m)^2 \, dS \; + \; k_G  \int_{\Gamma} G \, dS\, ,
\end{equation}
which indeed coincide with the Cahnam-Helfrich functional given as Eq.~(1) in the main text.

\begin{figure}
	\includegraphics[width=12cm]{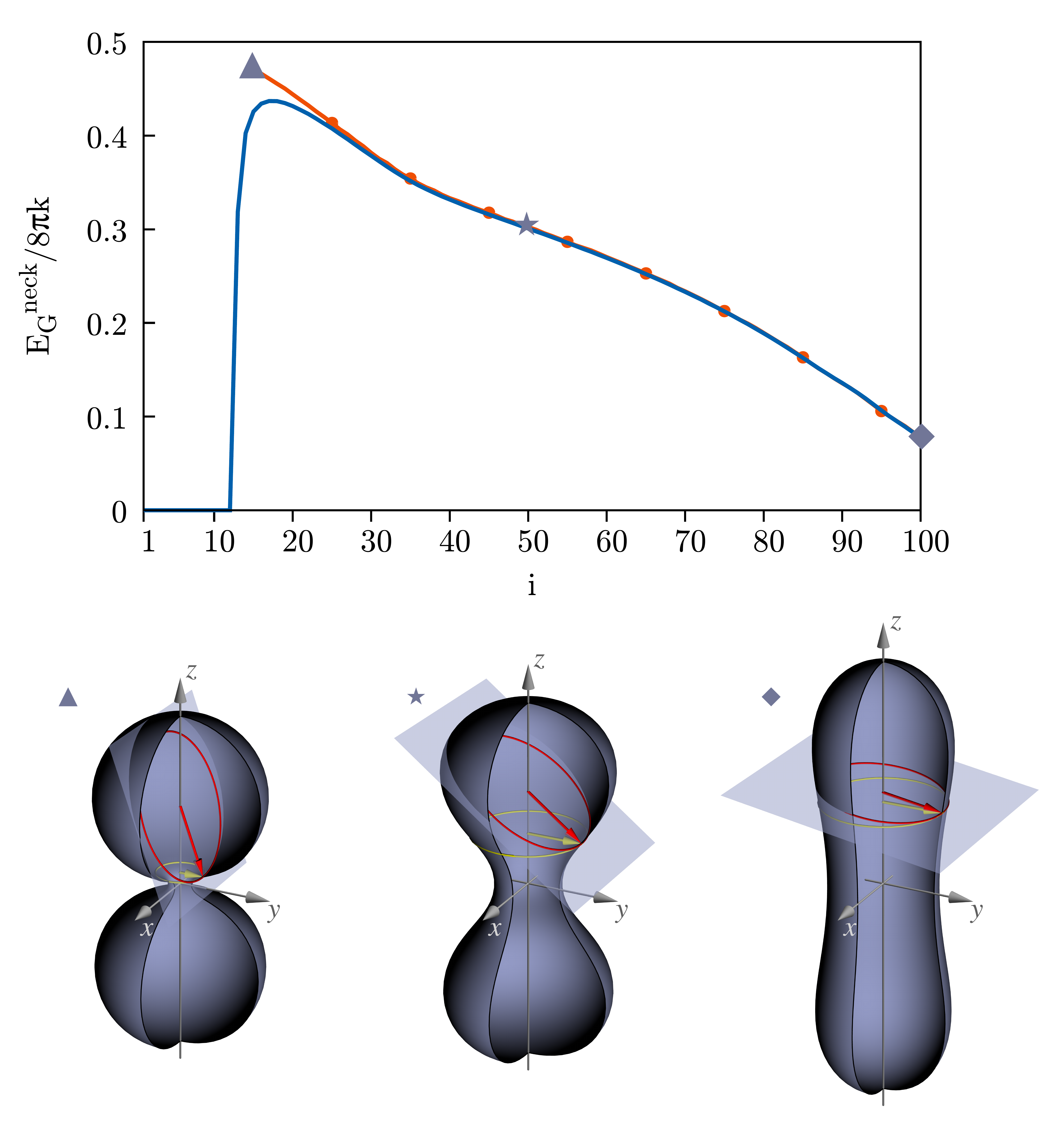}
	\caption{{\small Top panel: normalized Gaussian energy of the neck along the MEP (blue line). The orange line with dots provides the neck Gaussian energy as a post-processing based on the sharp interface Canham-Helfrich energy, computed considering the $\phi = 0$ isoline as the membrane midsurface: $E^{\text{neck}}_G(Z) / (8 \pi k) \approx \sqrt{1 - (r(Z)/R_n)^2}/2$. The agreement between the two curves progressively deteriorates when getting closer to the saddle point, due to the increasing curvature of the membrane generatrix. Bottom panel: three membrane configurations sketching the upper (yellow) circular boundary of the neck with its curvature radius $r(Z)$ (yellow arrow), and the osculating (red) circle to the vesicle cross section with the cutting plane passing through the neck boundary and containing both the surface normal and the tangent to the circle. The radius $R_n$ of the osculating circle is shown as a red arrow. The position of each configuration along the MEP is denoted by the corresponding symbol (triangle, star and rhombus).}}
	\label{Neck_energy_SM}
\end{figure}

From the above analysis, denoting $\phi_e(\bm{x},\lambda) = {\rm argmin} \, E[\phi, \lambda]$, it follows that the surface $\Gamma_e = \left\{\bm{x}: \lim\limits_{\lambda \rightarrow 0}\phi_e(\bm{x},\lambda) = 0 \right\}$ is a minimizer of the sharp interface energy whenever $\Gamma_e$ is sufficiently smooth to allow the evaluation of the Cahnam-Helfrich functional. 

The above reasoning can be extended to the case where the field is subject to the constraint on the vesicle areas and enclosed volumes, with the phase-field area and volume estimators given in Eqs.~(19) and (20) in the main text.

\section{Geometrical interpretation of the neck energy}\label{Neck_energy_geometry}

With reference to Fig. 4 of the main text (reported below for convenience), proceeding to the right along the MEP, beyond the saddle point, the computed Gaussian energy of the neck progressively decreases. In order to get a geometrical understanding of this decrease, for each image along the MEP, we extract the $\phi = 0$ isoline in the $r-z$ plane from the phase-field and find its interpolating polynomial $r = r(z)$. By resorting to the Gauss-Bonnet theorem,
\begin{equation}\label{Gauss_Bonnet_theorem}
\int_{\Gamma} G \; dS = 2 \pi \chi(\Gamma) - \int_{\partial \Gamma} k_g \; dl,
\end{equation}
the total curvature of the neck (a surface with Euler characteristic $\chi = 0$) is (minus) the integral of the geodesic curvature $k_g$ along the neck boundary (made up of two circles of radius $r(\pm Z)$, in the two planes normal to the z-axis delimiting the neck). The square of the geodesic curvature of these circles is $k_g^2 = k^2 - k_n^2$, where $k = 1 / r(\pm Z)$ and $k_n$, namely the normal curvature, is the curvature of the vesicle section with the plane containing both the surface normal and the tangent to the circles, evaluated at the neck boundary. The bottom panel of the figure shows a few vesicle configurations, with their upper neck boundary (yellow) and the osculating (red) circle defined by the radius of normal curvature
\begin{equation}\label{Neck_normal_curvature}
 \dfrac{1}{k_n} = r(\pm Z) \sqrt{1 + \dfrac{dr}{dz} \bigg|_{\pm Z}^2} \; .
\end{equation}
Approximating the vesicle with its sharp-interface counterpart, the neck Gaussian energy is  $E^{\text{neck}}_G(Z) / (8 \pi k) \approx r(\pm Z) k_g / 2 =  \sqrt{1 - (r/R_n)^2}/2$, where $R_n = 1/k_n$. From the insets of the figure, near the saddle point (configuration denoted by the triangle) $R_n >> r(\pm Z)$, leading to $k_g \approx 1/r(\pm Z)$. Moving away from the saddle point, the two curvature radii become eventually comparable, $R_n \approx r(\pm Z)$, leading to $k_g \approx 0$.  The main plot in the figure provides the actual Gaussian energy of the neck as computed from the phase-field (blue line) compared to the sharp-interface approximation (orange line with dots). The agreement is excellent as long as different membrane segments do not approach each other to a distance smaller than the bilayer thickness. Indeed, the agreement progressively deteriorates when getting closer to the saddle point, due to the increasing curvature of the membrane generatrix. Actually, one could better and better reproduce the sharp interface energy also in this stage by reducing the regularizing parameter $\lambda$. On the other hand, from a physical point of view, the thickness of the bilayer is finite, making the sharp interface model inappropriate when the saddle point is approached. Overall, these results confirm the accuracy of the proposed phase-field expression for the Gaussian curvature.


%